\documentclass[a4paper,12pt]{article}
\usepackage{latexsym,amsmath,amsfonts,amsbsy,amscd,amsxtra,amsgen,amsopn,bbm,amsthm,amssymb}

\begin{document}
\title{\textbf{Absolute quantum energy inequalities}}
\author{Christopher J. Fewster${}^{(1)}$\footnote{Electronic address: {\tt cjf3@york.ac.uk}}\, \& 
Calvin J. Smith${}^{(1,2)}$\footnote{Electronic address: {\tt calvin.smith@ucd.ie}} \\ \textit{${}^{(1)}$Department of Mathematics, University of York,}
\\ \textit{Heslington, York, YO10 5DD, UK}\\
\textit{${}^{(2)}$School of Mathematical Sciences, University College Dublin,} \\ \textit{Belfield, Dublin 4, Ireland}\footnote{Address from 1 January 2007.}}
\date{20 December 2007}

\maketitle
\newtheorem{theorem}{Theorem}[section]
\newtheorem*{defn}{Definition}
\newtheorem{lemma}[theorem]{Lemma}
\newtheorem{corollary}[theorem]{Corollary}
\newtheorem{claim}[theorem]{Claim}
\newtheorem{prop}[theorem]{Proposition}
\newtheorem{proposal}[theorem]{Proposal}
\newtheorem{example}[theorem]{Example}
\newtheorem{summary}[theorem]{Summary}
\newtheorem{axiom}{Wald Axiom}
\newcommand{\supp}{\textrm{supp }}
\newcommand{\singsupp}{\textrm{singsupp }}
\newcommand{\QED}{\qed \\}
\newcommand{\R}{\mathbb{R}}
\newcommand{\Z}{\mathbb{Z}}
\newcommand{\C}{\mathbb{C}}
\newcommand{\N}{\mathbb{N}}
\newcommand{\M}{\mathcal{M}}
\newcommand{\D}{\mathcal{D}}
\newcommand{\U}{\mathcal{U}}
\newcommand{\E}{\mathcal{E}}
\newcommand{\s}{\mathcal{S}}
\newcommand{\open}{\mathcal{O}}
\newcommand{\mink}{\mathbb{M}}
\newcommand{\norm}{\parallel}
\newcommand{\hilbert}{\mathcal{H}}
\newcommand{\vol}{\mathrm{vol}}
\newcommand{\diag}{\mathrm{diag\,}}
\newcommand{\Char}{\mathrm{Char \,}}
\newcommand{\LIM}{\mathrm{l.i.m.}}
\newcommand{\loc}{\mathrm{loc}}
\newcommand{\nnnl}{\nonumber \\}
\newcommand{\ren}{\mathrm{ren}}
\newcommand{\normal}{\mathcal{N}}
\newcommand{\GL}{\mathrm{GL}_n(\C)}
\newcommand{\dvol}{\mathrm{dvol}}
\newcommand{\rmd}{\mathrm{d}}
\newcommand{\tr}{\mathrm{tr }}

\begin{abstract}
Quantum Energy Inequalities (QEIs) are results which limit
the extent to which the smeared renormalised energy density of the quantum
field can be negative, when averaged along a
timelike curve or over a more general timelike submanifold in
spacetime. On globally hyperbolic spacetimes the minimally-coupled
massive quantum Klein--Gordon field is known to obey a `difference' QEI
that depends on a reference state chosen arbitrarily from the class of Hadamard states.
In many spacetimes of interest this bound cannot be evaluated explicitly. In this paper we obtain the first
`absolute' QEI for the
minimally-coupled massive quantum Klein--Gordon field on four dimensional
globally hyperbolic spacetimes; that is, a bound which depends only on the local geometry.
The argument is an adaptation of that used to prove the
difference QEI and utilises the Sobolev wave-front set to give a complete characterisation of the singularities of
the Hadamard series. Moreover, the bound is explicit and can be
formulated covariantly under additional (general) conditions. We also
generalise our results to incorporate adiabatic states. 
\end{abstract}

\noindent\emph{Dedicated to Klaus Fredenhagen on the occasion of his 60th birthday.}

\section{Introduction}

The classical minimally coupled scalar field, like most matter models
studied in classical general relativity, obeys the weak energy condition
(WEC). That is, the stress energy tensor $T_{ab}$ obeys the inequality
$T_{ab}v^av^b\geq 0$ for all timelike vector fields $v^a$, which entails
that observers encounter only non-negative energy densities. However, it has been
known since 1965 that no Wightman quantum field theory can
obey the weak energy condition \cite{epstein}, (see \cite{fewster 
lisbon, fewster paris}
for simple arguments as to why this is true). Moreover, under many
circumstances there is no lower bound to the energy densities available
in quantum field theory (QFT). This surprising feature of QFT has often
been used to support proposals for exotic spacetimes, such as warp
drive and traversable worm holes, which require WEC-violating matter
distributions. In addition, the validity of the second law of
thermodynamics is called into question by WEC violations in QFT
\cite{ford}.

With these concerns in mind, substantial effort has been directed to
understand the magnitude and extent of negative energy densities
permitted by QFT, starting with the work of Ford in
1978 \cite{ford}. Given the failure of pointwise energy inequalities,
attention has been focussed on averages of the stress tensor along
timelike worldlines or over spacetime regions. In many QFT models such
averages turn out to obey Quantum Energy Inequalities (QEIs); that is,
their expectation values are bounded from below as the state varies
within the class of physically reasonable states \cite{fewster eveson,
fewster, fewster teo, fewster pfenning, ford, ford and roman, ford and
pfenning, pfenning}. Since their inception QEIs
have been applied to a variety of physical problems; they form the basis
of the arguments constraining exotic spacetimes such
as the warp drive \cite{alcubierre, ford and pfenning} and
traversable wormholes \cite{ford and roman, fewster roman}. 

In their most common form, QEIs place lower bounds on the expectation value
of the averaged stress energy tensor relative to that obtained in a
reference state. For this reason they are called \textit{difference QEIs}. To
give a specific example \cite{fewster}, consider the minimally coupled scalar field of
mass $\mu \geq 0$ in a globally hyperbolic spacetime $(\M,g)$ of
dimension $m\geq 2$, and let $\gamma : \R\to \M$ be a smooth
timelike curve (not necessarily a geodesic) with velocity $v^a$. Then
for any real-valued $f\in C^\infty_0(\R)$ the QEI 
\begin{equation}\label{gwlqi}
\int_\R \rmd t \, f^2(t) \, \bigg(\langle v^a v^b T^\mathrm{ren}_{ab}
\rangle_{\omega}(\gamma(t)) - \langle v^a v^b T^\mathrm{ren}_{ab} \rangle_{\omega_0}(\gamma(t))  
\bigg)  \geq - \mathcal{B}_D 
\end{equation}
holds for all Hadamard states $\omega$, where the bound 
\begin{equation}\label{gwlqib}
\mathcal{B}_D = \int^\infty_0 \frac{\rmd\xi}{\pi}\bigg[ f\otimes f \,
\vartheta^*\langle v^a v^{b^\prime} T^\mathrm{split}_{ab^\prime} \rangle_{\omega_0} 
\bigg]^\wedge (-\xi,\xi)
\end{equation}
depends only on $f$, $\gamma$, and the reference state $\omega_0$ (which
may be any Hadamard state); note that it does not depend on
the state of interest $\omega$. Here the hat denotes the Fourier
transform given, in our conventions, by $\widehat{f}(\xi)=\int_{\R^n}\rmd^n x\, f(x)e^{i\xi\cdot
x}$. The quantity $\langle v^a v^{b^\prime} T^\mathrm{split}_{ab^\prime}
\rangle_{\omega_0}(x,x^\prime)$ is the (unrenormalised) point split energy
density defined, in a neighbourhood of $\gamma$, by
\begin{equation}
\langle v^a v^{b'} T^\mathrm{split}_{ab'} \rangle_{\omega_0}(x,x') = \bigg(
\frac{1}{2}\sum_{\alpha=0}^3 e^a_\alpha \nabla_a\otimes
e^{b'}_\alpha\nabla_{b'}+\frac{1}{2}\mu^2\mathbbm{1}\otimes\mathbbm{1}\bigg) \Lambda_{\omega_0}(x,x')
\end{equation}
where $\{ e^a_\alpha \}_{\alpha=0,1,2,3}$ is a tetrad field satisfying
$e^a_0|_\gamma =v^a$ (see Section 3 of \cite{fewster} for a more
detailed discussion) and $\Lambda_{\omega_0}$ is the two point function
of the state $\omega_0$. Finally, $\vartheta^*\langle v^a v^{b^\prime} T^\mathrm{split}_{ab'}\rangle_{\omega_0}$ denotes the
pull-back 
\begin{equation}
\vartheta^*\langle v^a v^{b^\prime} T^\mathrm{split}_{ab^\prime}\rangle_\omega (\tau,\tau^\prime)= \langle
v^a v^{b^\prime}T^\mathrm{split}_{ab^\prime} \rangle_{\omega_0}(\gamma(\tau),\gamma(\tau^\prime))
\end{equation}
which may be defined rigorously as a distribution on $\R^2$ using the
techniques of microlocal analysis, which also guarantee
that the bound (\ref{gwlqib}) is finite. Similar bounds also hold for
the free spin-$1/2$ and spin-$1$ field in comparable generality and
rigour \cite{dawson, fewster verch, fewster mistry, fewster pfenning}.
Recently, quantum energy inequalities have also been proven for free spin-3/2
fields in Minkowski spacetime \cite{yu wu, hu ling zhang}.

As already mentioned, one application of the QEIs is to place constraints
on exotic spacetimes. However, the bound given above, while valid in any
globally hyperbolic curved spacetime, depends crucially on the choice of
a reference state. Although Hadamard states
exist on any globally hyperbolic spacetimes \cite{fulling sweeny wald, fulling narcowich wald}, closed form expressions for
two point functions are known only in very special circumstances. For
instance, no such expression is available for any Hadamard state on the
warp drive spacetime. Typically these problems have been avoided by
heuristic appeals to the equivalence principle to justify the use of Minkowski spacetime QEIs
on sufficiently small scales. To date, this approach, while physically
reasonable, lacks full mathematical justification and control over the
scales on which it is valid. It would clearly be preferable to employ a lower
bound which did not require the specification of a reference state and
placed constraints directly on $\langle T_{ab}^\ren \rangle_\omega$.
Bounds of this type, known as \textit{absolute QEIs}, have been established in
flat spacetimes \cite{flanagan1} but the only curved spacetime absolute QEI
is that of Flanagan \cite{flanagan2} (see also \cite{fewster 2d, vollick}) which applies to massless free fields in two-dimensional
globally hyperbolic spacetimes. This approach relies on the
conformal invariance of the theory and does not generalise to higher
dimensions or non-zero mass. However, it does provide the basis for a
QEI on arbitrary positive energy conformal field theories in Minkowski
spacetime, including interacting examples \cite{fewster hollands}.

In this paper we present the first absolute QEI applicable to the scalar
field of mass $\mu\geq 0$ in four-dimensions, by refining and modifying the argument presented in \cite{fewster}. For
averaging along
timelike worldline $\gamma$, our result takes the form
\begin{equation}
\int_\R \rmd t \, f^2(t) \langle v^a v^b T^\ren_{ab} \rangle_\omega(t) \geq - \mathcal{B}_A\,,
\end{equation}
where 
\begin{equation}
\mathcal{B}_A = \int_{\R^+} \frac{\rmd\xi}{\pi} \bigg[ f\otimes f \, \vartheta^*T^\mathrm{split}\widetilde{H}   \bigg]^\wedge
(-\xi,\xi) + \text{``local curvature terms"}
\end{equation}
and $T^\mathrm{split}\widetilde{H}$ is constructed in the same fashion as
$\langle v^a v^{b^\prime}T^\mathrm{split}_{ab^\prime}\rangle_{\omega_0}$ but with (essentially) the
first few terms of the Hadamard series replacing the reference two-point function. At
the technical level, we invoke a refined version of microlocal
analysis which keeps track of the order of singularities.

The structure of this paper is as follows. In Section \ref{QFTCST} we review
the algebraic formulation of quantum field
theory in curved spacetimes, review two (equivalent) formulations of the Hadamard
condition and give a detailed analysis of the singularity structure of
the Hadamard series in terms of Sobolev wave-front sets.
Section \ref{QEI section} contains our main result, theorem \ref{mgi},
which is then used to give a number of examples of absolute QEIs.
Although our bound depends on a choice of coordinates, we describe how
the dependence can be eliminated by restricting the choice of smearing
tensor, thus providing a covariant formulation of our bounds.

\section{Quantum field theory in curved spacetime}\label{QFTCST}

\subsection{The algebra of observables and the first definition of \\
Hadamard states}\label{qftcst1}

We shall employ the algebraic framework for describing the scalar quantum
field in a classical curved four-dimensional spacetime $(\mathcal{M},g)$.
Here $\mathcal{M}$ is a four-dimensional smooth manifold (assumed Hausdorff,
paracompact and without boundary) with a Lorentz metric $g_{ab}$ of signature
($+---$). Furthermore, we require $(\M,g)$ to be \textit{globally hyperbolic}, that is
$\M$ contains a Cauchy surface. Where index notation is
used, Latin indices will run over the range $0,1,2,3$ unless explicitly
stated otherwise, while Greek characters will denote frame indices and
also run over $0,1,2,3$ unless explicitly stated otherwise. We employ units
in which $c=\hbar=1$. 

The minimally coupled scalar field $\phi$ obeys the Klein--Gordon equation $(\nabla^2 + \mu^2)\phi
=~0$,  where $\nabla^2 = g^{ab}\nabla_a \nabla_b$ and $\mu
\geq 0$ is the mass of the field quanta. Global hyperbolicity entails the
existence of unique global advanced $(E^-)$ and retarded
$(E^+)$ Green functions $E^\pm : C^\infty_0 (\M)\to C^\infty(\M)$
for the Klein--Gordon equation obeying
\begin{equation}
(\nabla^2+\mu^2)E^\pm f = E^\pm (\nabla^2+\mu^2)f =f , 
\end{equation}
and
\begin{equation}
 \supp
E^\pm f \subset J^\pm(\supp f) 
\end{equation}
for all $f\in C^\infty_0(\M)$, where $J^\pm(S)$ denote the causal future
($+$) and past ($-$) of a set $S$. One may use the set of smooth
functions having compact support in $\M$, $C^\infty_0(\M)$, to label a set of abstract objects $\{ \phi(f)
\mid f \in C^\infty_0 (\M) \}$ which generate a free unital $*$-algebra
$\mathfrak{A}$ over $\C$. The \textit{algebra of smeared fields}
$\mathfrak{A}(\M,g)$ is defined to be the quotient of 
$\mathfrak{A}$ by the following relations:  \\
\begin{tabular}{cl}
i) & Hermiticity, $\phi(f)^* = \phi(\overline{f})$ $\forall f \in
C^\infty_0 (\M)$; \\
ii) & Linearity, $\phi(\alpha f + \beta f^\prime)=\alpha \phi(f)+\beta\phi(f^\prime)$ $\forall \alpha,\beta \in \C$ and $\forall f,f^\prime \in C^\infty_0(\M)$;
\\
iii) & Field equation, $\phi((\nabla^2+\mu^2)f)=0$ $\forall f \in
C^\infty_0 (\M)$; \\
iv) & Canonical commutation relations,
$[\phi(f),\phi(f^\prime)]=iE(f, f^\prime)\mathbbm{1}$ $\forall f,f' \in
C^\infty_0 (\M)$.
\end{tabular} \\
Here, $E=E^--E^+$ is the advanced-minus-retarded Green's function for the Klein--Gordon
operator and by $E(f,f^\prime)$ we mean 
\begin{equation}
E(f , f^\prime) = \int_\M \dvol(x) \, f(x)(Ef^\prime)(x) \, .
\end{equation}
It is relation (iv) that quantises the
field theory.

In this framework, a \textit{state} is a linear functional $\omega$ on $\mathfrak{A}(\M,g)$ which is
normalised so that $\omega(\mathbbm{1})=1$ and is positive in the
sense that $\omega(A^*A)\geq0$ for all $A\in\mathfrak{A}(\M,g)$. The two point function associated with the state $\omega$ is a bilinear
map $\Lambda_\omega : C^\infty_0 (\M) \otimes
C^\infty_0 (\M) \to \C$ given by $\Lambda_\omega (f , f^\prime) =
\omega(\phi(f)\phi(f^\prime))$. We will only consider states for which
$\Lambda_\omega$ is a distribution, i.e., $\Lambda_\omega \in \D^\prime
(\M\times\M)$. It is clear from (iv) that the antisymmetric part of
$\Lambda_\omega$, 
\begin{equation}\label{anti part state indep}
\frac{1}{2}\bigg( \Lambda_\omega (f,f^\prime)-\Lambda_\omega(f^\prime,f) 
\bigg)  =\frac{i}{2}E(f,f^\prime) \, ,
\end{equation}
is state independent. 

As already mentioned, we will largely be concerned with Hadamard states. There are two
equivalent formulations of the Hadamard condition, both
of which will be used in the sequel. The original definition, given in
a precise form by Kay \& Wald \cite{kay}, involves a local series expansion of
the two point function $\Lambda_\omega$ associated with a state $\omega$
and is based upon Hadmard's work on the fundamental solution for
hyperbolic operators.

In order to give the precise formulation of the Hadamard series
construction we first introduce some geometrical structures, following
\cite{kay,radzikowski,sahlmann2}. We denote by $\mathfrak{X}\subset \M\times\M$ the
set
\begin{eqnarray}
\mathfrak{X} &=& \{ (x,x^\prime)\in\M\times\M \mid  x,x^\prime \text{ causally
related and} \nnnl &\,& \qquad J^+(x)\cap J^-(x^\prime) \,\&\, J^-(x)\cap J^+(x^\prime)
\text{ are
contained} \nnnl &\,& \qquad \text{within a convex normal neighbourhood} 
\} \, .
\end{eqnarray}
For each $(x,x')\in \mathfrak{X}$ let $U_{x,x'}$ be any convex normal
neighbourhood containing $J^+(x)\cap~ J^-(x')$ and $J^-(x)\cap J^+(x')$. 
Then (cf. Lemma 3.1 in~\cite{radzikowski}) $X=\bigcup_{(x,x')\in\mathfrak{X}}
U_{x,x'}\times U_{x,x'}$ is an open neighbourhood of $\mathfrak{X}$ in $\M\times \M$ on
which the signed\footnote{We adopt the convention that $\sigma(x,x')>0$ if
$x,x'$ are spacelike separated, $\sigma(x,x')<0$ if $x,x'$ are timelike separated
and $\sigma(x,x')=0$ if they are null separated. In Minkowski spacetime, $(\R^4,\eta)$,
for example $\sigma(x,x')=-\eta_{ab}(x-x')^a(x-x')^b$.} squared geodesic separation of
points $\sigma$ is well-defined and smooth, and on which the Hadamard construction
(to be described shortly) can be carried out. 
Any open neighbourhood of $\mathfrak{X}$ defined in this way will be called a \textit{regular
domain}.

For each $k=0,1,2,\dots,$ we may define a distribution $H_k \in
\D^\prime (X)$ by 
\begin{eqnarray}\label{parametrix}
H_k(x,x^\prime) &=&  
\frac{1}{4\pi^2} \bigg\{ \frac{\Delta^\frac{1}{2}(x,x^\prime)}{\sigma_+ (x,x^\prime)}
+ \sum^k_{j=0}v_j (x,x^\prime)\frac{\sigma^j(x,x^\prime)}{\ell^{2(j+1)}}
\ln\bigg(\frac{\sigma_+ (x,x^\prime)}{\ell^2}\bigg)
\nonumber \\
&\,& \qquad 
+\sum^k_{j=0}w_j(x,x^\prime)\frac{\sigma^j(x,x^\prime)}{\ell^{2(j+1)}}\bigg\}
\, ,
\end{eqnarray}
where we have introduced a length scale $\ell$ to make
$\sigma/\ell^2$ dimensionless\footnote{A different choice of length
scale $\ell'$ can be absorbed into a redefinition of the local curvature terms
$C_{ab}$ appearing in the renormalisation of the stress-energy tensor; see
Section \ref{ren}.} and the coefficient functions $\Delta$, $v_j$ and $w_j$ will be
explained below. We also set $H_{-1}=\Delta^{1/2}/(4\pi^2\sigma_+)$. By $F(\sigma_+)$, for some
function $F$, we mean
the distributional limit
\begin{equation}
F(\sigma_+) = \lim_{\epsilon\rightarrow 0^+}F(\sigma_\epsilon)  \, ,
\end{equation}
where $\sigma_\epsilon(x,x^\prime) =
\sigma(x,x^\prime)+2i\epsilon(t(x)-t(x^\prime))+\epsilon^2$ and $t$ is a
time function; that is, $\nabla^a t$ is a normalised future directed timelike vector
field on $X$. We shall occasionally use the notation $t=t(x)$
and $t^\prime=t(x^\prime)$. The function $\Delta \in C^\infty (X)$
is the van Vleck-Morette determinant bi-scalar and is given by
\begin{equation}
\Delta (x,x^\prime) = -\frac{\det\big(-\nabla_a\otimes\nabla_{b^\prime}\,\,\sigma(x,x^\prime)\big)
}{\sqrt{-g(x)}\sqrt{-g(x^\prime)}} \, .
\end{equation}
The functions $v_j$ and $w_j$ are found by fixing $x^\prime$
and applying $(\nabla^2 + \mu^2)\otimes\mathbbm{1}$ to $H_k$ and equating all the
coefficients of $1/\sigma_+$, $1/\sigma^2_+$, $\ln \sigma_+$ etc to zero. This
determines a system of equations (known as the Hadamard recursion
relations, given in appendix \ref{hadamard recursion relations}) which can be solved uniquely (in $X$) for the
$v_j$ series. The $w_j$ series is specified once the value of $w_0$ is
fixed; we adopt Wald's prescription that $w_0=0$ \cite{wald}. We remark
that the $k\to\infty$ limit of the right-hand side of (\ref{parametrix})
has a nonzero radius of convergence in analytic spacetimes, but not in general \cite{gunther}. 

Let $\normal$ be a causal normal neighbourhood of a Cauchy surface
$\mathcal{C}$ \cite{kay}; that is, $\mathcal{C}$ is a Cauchy surface for $\normal$
and every double-cone $J^+(x)\cap J^-(y)$ with $x,y\in\normal$ is
contained in a convex normal neighbourhood of $(\M,g)$ (see Lemma 2.2
in~\cite{kay} for the existence of causal normal neighbourhoods). We may further
choose an open neighbourhood $X_*$ of the set of pairs
of causally related points in $\normal\times\normal$ whose closure is contained in $X\cap
(\normal\times\normal)$ and a
cut-off function $\chi:\mathcal{N}\times\mathcal{N}\to [0,1]$ so that
\begin{equation}
\chi |_{X_*} =1 \quad \text{and} \quad
\chi |_{(\mathcal{N}\times\mathcal{N})\setminus X}=0 \, .
\end{equation}
See Lemma 3.3 in \cite{radzikowski} for the existence of $X_*$ and
$\chi$ with these properties. 

Given the above, a state $\omega$ on $\mathfrak{A}(\M,g)$ is said to be \textit{Hadamard}
if for each $k\in\N$ there exists a $F_k\in C^k(\mathcal{N}\times\mathcal{N})$ such
that 
\begin{equation}\label{hadamard2}
\Lambda_\omega = \chi H_k + F_k \, 
\end{equation}
in $\normal\times\normal$. We remark that this definition can be shown
to be independent of the choices of $\mathcal{C}$, $\normal$, $t$, $\chi$, $X$ and
$X_*$ \cite{kay,sahlmann2}. 

In the special case in which $\M$ is a convex normal neighbourhood, we
note that $\M$ would be a causal normal
neighbourhood of any of its Cauchy surfaces and we could take
$X_*=X=\M\times\M$ and $\chi\equiv 1$, so \eqref{hadamard2}
becomes $\Lambda_\omega=H_k +F_k$ and holds on the whole of
$\M\times\M$. In the general case, it is easy to see (e.g.,  using the
microlocal characterisation of the Hadamard condition)
that if $\omega$ is Hadamard then so is its restriction to any open
globally hyperbolic subset of $\M$, considered as a spacetime in its own
right. Thus $\Lambda_\omega-H_k$ is $C^k$ for all $k$ on any set of the form
$U\times U$ where $U$ is a globally hyperbolic convex normal neighbourhood.
As every point $x\in\M$ has such a neighbourhood $U_x$ we may conclude
that $\Lambda_\omega-H_k$ is $C^k$ for all $k$ in an open neighbourhood of the diagonal of the form
$\bigcup_{x\in\M} U_x\times U_x$; we will refer to any
such open neighbourhood as an {\em ultra-regular domain}.  

{\noindent\em Note:} The need to introduce the notion of an
ultra-regular domain only came to light as the final version of this paper was prepared for
publication, and after \cite{smith} had gone to press. 
One of us (CJS) would like to warn the reader that some results in
\cite{smith} hold on ultra-regular domains as opposed to the stated regular
domain; with this modification, the results of \cite{smith} are unchanged.

\subsection{Renormalisation of the stress tensor}\label{ren}

The Hadamard series construction forms the basis for the renormalisation of the
stress-energy tensor in curved spacetimes, to which we
now turn. The classical stress tensor of the real scalar field
\begin{equation}
T_{ab}(x) = \bigg(\nabla_a \otimes \nabla_b -
\frac{1}{2}g_{ab}g^{cd}\nabla_c\otimes\nabla_d + \frac{1}{2}\mu^2g_{ab}\mathbbm{1}\otimes\mathbbm{1} 
\bigg)\big( \varphi\otimes\varphi\big) (x,x)
\end{equation}
must be renormalised in QFT owing to the divergent behaviour of the two
point function. Define the \textit{point-split stress-energy operator}
(which should not be confused with the stress-energy tensor itself) by
\begin{equation}\label{pt split op}
T^\mathrm{split}_{ab^\prime} = \nabla_a \otimes \nabla_{b^\prime} -
\frac{1}{2}g_{ab^\prime}g^{cd^\prime}\nabla_c\otimes\nabla_{d^\prime} + \frac{1}{2}\mu^2g_{ab^\prime} \mathbbm{1}\otimes\mathbbm{1}
\end{equation}
near the diagonal in $\M\times\M$, where $g_{ab^\prime}(x,x')$ is the parallel
propagator\footnote{Thus, if $v^{b'}\in T_{x'}\M$, $g_{ab'}(x,x')v^{b'}$
is its parallel transport to $T_{x}\M$ along the unique geodesic joining
$x'$ to $x$, which is well-defined sufficiently close to the diagonal in $\M\times\M$.}.
If $\omega$ is a
Hadamard state we may define $\langle T_{ab}^\ren\rangle_\omega(x)$ at any point
$x\in\M$ by the following procedure: \\
\noindent 
a) note that $\Lambda_\omega - H_k \in C^2(X)$ for $k\geq 2$ and any
ultra-regular domain $X$, so $T^\mathrm{split}_{ab^\prime}\big( \Lambda_\omega - H_k  \big)$ is
defined and continuous near the diagonal in $\M\times\M$; \\
b) define 
\begin{equation}\langle T^\mathrm{fin}_{ab}\rangle_\omega (x)=
\lim_{x^\prime\rightarrow x}g_b^{\,\,\,
b^\prime}(x,x^\prime)T^\mathrm{split}_{ab^\prime}\big(\Lambda_\omega - H_k 
\big)(x,x^\prime)
\end{equation}
for $k\geq 2$;\\
c) make finite corrections to $\langle
T^\mathrm{fin}_{ab}\rangle_\omega$ in order to obtain a conserved tensor
$\langle T_{ab}^\ren\rangle_\omega(x)$ with the correct properties in Minkowksi space. 

Step (c) is needed because the tensor $\langle T^\mathrm{fin}_{ab}\rangle_\omega$ is not
covariantly conserved and cannot be considered as an appropriate
stress-energy tensor (it could not be inserted on the right hand side of the Einstein
equations, for example). However, it turns out that
$\nabla^a\langle T^\mathrm{fin}_{ab}\rangle_\omega$ is of the form
$\nabla_b Q$ where $Q$ is a local quantity, determined up to a constant
\cite{wald}; subtracting $Qg_{ab}$ by hand from $\langle
T^\mathrm{fin}_{ab}\rangle_\omega$ we therefore obtain a conserved quantity. The
undetermined constant in $Q$ is fixed by the
requirement that in Minkowski spacetime the vacuum
expectation value vanishes. If we require that the difference $\langle
T^\mathrm{ren}_{ab}\rangle_\omega- \langle T^\mathrm{ren}_{ab}\rangle_{\omega_0}$ should be given by 
\begin{equation}
\langle T^\mathrm{ren}_{ab}\rangle_\omega- \langle T^\mathrm{ren}_{ab}\rangle_{\omega_0} =
\lim_{x^\prime\rightarrow x}g_b^{\,\,\, b^\prime}(x,x^\prime)
T^\mathrm{split}_{ab^\prime}\big(\Lambda_\omega - \Lambda_{\omega_0}\big)(x,x^\prime)
\end{equation}
then any remaining finite renormalisation must
take the form of a state-independent conserved local curvature
term $C_{ab}$ that vanishes in Minkowski space, and the finite renormalised expection value of the
quantum stress energy is given by
\begin{equation}\label{Tren}
\langle T^\mathrm{ren}_{ab}\rangle_\omega(x) = \langle
T^\mathrm{fin}_{ab} \rangle_\omega(x) - Q(x)g_{ab}(x) + C_{ab}(x) \, .
\end{equation}
We take the view that the tensor $C_{ab}$ is a necessary part of the specification of a given species of scalar
field, alongside the mass and curvature coupling. Given sufficient
experimental accuracy $C_{ab}$ should, in principle, be measurable.

The renormalisation prescription we have outlined
above is vulnerable to the criticism that the $Qg_{ab}$ term needed to
restore conservation of $\langle T^\ren_{ab}\rangle_\omega$
is only found to be a local curvature term {\em a posteriori}. Moretti \cite{moretti}
has shown that this problem can be circumvented by an alternative construction of the quantum stress
tensor. The basic idea is to modify the classical stress energy tensor $T_{ab}$ by adding a term of the form
$\alpha\varphi(\nabla^2+\mu^2)\varphi$ for constant $\alpha$. While this addition does not
affect the classical physics, it has a non-trivial quantisation. A judicious choice of
$\alpha$ ensures that the quantised
stress energy tensor is conserved {\em a priori} (see theorem 2.1 of
\cite{moretti}) and agrees with the usual quantisation up to conserved
local curvature terms. Although Moretti's approach is certainly elegant,
it turns out that the usual quantisation is better adapted to the
derivation of QEIs. In particular, our argument relies crucially on being able to write
$T^\mathrm{split}_{ab'}$ in a symmetric form which is not possible with
a term of the form $\mathbbm{1}\otimes (\nabla^2+\mu^2)$. A similar
problem arises for the non-minimally coupled scalar field. In this case
one must smear the stress-energy tensor even to obtain an inequality on
the classical field \cite{fewster osterbrink}, which necessitates a more
complicated analysis at the quantum level \cite{fewster osterbrink II}.

\subsection{The wave-front set and second definition of Hadamard state}

The above discussion shows that Hadamard states are characterised by
their singularity structure. For this reason the techniques of
microlocal analysis, which focus attention on singular behaviour, are
ideally suited to this theory. This realisation has led to a number of
important developments in the theory of quantum fields in curved
backgrounds, following initial work of Radzikowski~\cite{radzikowski}, 
particularly in regard to renormalisation \cite{brunetti
fredenhagen,hollands wald a, hollands wald b}. In addition, 
the theory of (smooth) wave-front sets is a key tool in the proof of general difference
QEIs~\cite{fewster,fewster verch,fewster pfenning,dawson fewster} in general globally
hyperbolic spacetimes. Our absolute QEIs require the finer control on
singularities of distributions afforded by the Sobolev wave-front set.
In this subsection we briefly review the
definition of the smooth and Sobolev wave-front sets and explain how
they may be used to give a purely microlocal definition of the Hadamard condition, as first identified by
Radzikowksi~\cite{radzikowski}. In addition, we will state a result of
Junker and Schrohe \cite{junker} on the Sobolev wave-front set of the
two-point functions of Hadamard states. This will form the basis of our
analysis of the Sobolev wave-front sets of individual terms in the
Hadamard series in the next subsection. 

To begin, let $u \in \D^\prime (\R^m)$ be any distribution. We say that $u$ is
smooth at $x^\prime$ if there exists
an open neighbourhood $\open\subset\R^m$ of $x^\prime$ and a smooth function $\varphi\in
C^\infty(\open)$ such that $u(f)=\int_{\R^m}\rmd^m x \, \varphi(x)f(x)$
for all test functions $f\in C^\infty_0(\open)$. The \textit{singular support}, $\singsupp u$, of a distribution
$u\in\D^\prime(\R^m)$ is the complement in $\R^m $ of the set of all points at
which $u$ is smooth. In particular, a distribution is smooth
if and only if its singular support is empty.

While the singular support tells us `where' a distribution $u$ fails to be
smooth, Fourier transforms of localisations of $u$ contain
additional information. A covector $\zeta\in \R^m\setminus \{ 0\}$ is a \textit{direction of rapid decay 
for $u$ at $x$} if there exists a conic neighbourhood $\Gamma \subset \R^m\setminus
\{ 0 \}$ of $\zeta$ and a localiser $\chi \in C^\infty_0(\R^m)$
which does not vanish at $x$ such that
\begin{equation}
(1+|\xi|)^N |\widehat{\chi u}(\xi)|
\longrightarrow 0 \quad{\rm as}~\xi\to\infty~{\rm in}~\Gamma,\quad\forall N\in\N \, .
\end{equation}
The \textit{set of singular directions of $u\in\D^\prime(\R^m)$ at $x$}, $\Sigma_x(u)$, is
the complement in $\R^m\setminus\{ 0 \}$ of the set of directions of
rapid decay of $u$ at $x$. The wave-front set of $u$ assembles this information
in a convenient way (see Section 8.1 of \cite{hormander} for more detail).

\begin{defn}
The (smooth) wave-front set $WF(u)$ of a distribution $u\in\D^\prime(\R^m)$ is 
\begin{equation}
WF(u) = \{ (x,\xi)\in \R^m \times (\R^m\setminus\{ 0\}) \mid  \xi \in \Sigma_x(u)  \}
\end{equation}
\end{defn}

As an example, it is easy to verify that the Dirac $\delta$ and
Heaviside $\theta$ distributions have the following wave-front sets.
\begin{equation}\label{verify1}
WF(\delta)=WF(\theta)=\{ (0,\xi)
\mid \xi \in \R\setminus\{ 0\} \} \, .
\end{equation}
The wave-front set is a closed cone in $\R^m\times (\R^m\setminus \{ 0 \}
)$, whose elements transform as covectors under coordinate
transformations (see Theorem~8.2.4 in \cite{hormander}). Accordingly, the definition of wave-front set may
be extended to distributions on a smooth manifold $\M$ in the following way: We say $(x,\xi)\in
WF(u)\subset T^*\M\setminus \{ 0 \}$ if and only if there exists a chart neighbourhood $(\kappa,\U)$
of $x$ such that the corresponding coordinate expression of $(x,\xi)$ belongs
to $WF(u\circ\kappa^{-1})\subset \R^m\times (\R^m\setminus \{ 0 \} )$,
where $m$ is the dimension of $\M$.

The Sobolev wave-front set provides greater structure on the information in the
wave-front set. Recall that the Sobolev space $H^s(\R^m)$, $s \in \R$, is the set of all tempered
distributions $u$ on $\R^m$ such that
\begin{equation}
\int_{\R^m} \rmd^m\xi \, (1+|\xi|^2)^s |\widehat{u}(\xi)|^2 < \infty \, .
\end{equation}
We summarise some relevant properties of Sobolev spaces for convenience. 

\begin{prop}\label{Sobolev prop}
The Sobolev spaces $H^s(\R^m)$ have the following properties:
\begin{equation*}
\begin{array}{cl}
i) & \textrm{If $s>k+m/2$ then } H^s(\R^m) \subset C^k(\R^m) \text{ for } k\in\N;\\
ii) & H^s(\R^m) \subset H^{s^\prime}(\R^m) \, \forall s\geq s^\prime; \\
iii) & \textit{$H^s(\R^m)$ is closed under multiplication by smooth functions.}
\end{array}
\end{equation*}
\end{prop}

Associated with the scale of Sobolev spaces, there is a refined notion
of the wave-front set. Just as $WF(u)$ informs us where a
distribution $u$ fails to be smooth, the Sobolev wave-front set $WF^s(u)$ contains
information about where in phase space the distribution fails to be $H^s$. 

\begin{defn}
The distribution $u \in \D^\prime (\R^m)$ is said to be {\em microlocally $H^s$ at}
$(x,\xi)\in\R^m\times(\R^m\setminus\{0\})$ if there exists a
conic neighbourhood $\Gamma$ of $\xi$ and a smooth function $\varphi \in C^\infty_0
(\R^m)$, $\varphi(x)\not=0$, such that 
\begin{equation}
\int_\Gamma \rmd^m\zeta \, (1+|\zeta|^2)^s |[\varphi u]^\wedge(\zeta)|^2 < \infty \, .
\end{equation}
The Sobolev wave-front set $WF^s(u)$ of a distribution $u \in \D^\prime (\R^m)$ is the
complement, in $T^*\R^m\setminus \{0\}$, of the set of all pairs $(x,\xi)$ at which $u$
is microlocally $H^s$.
\end{defn}

It is easy to verify, for example, that 
\begin{equation}
WF^{s+1}(\theta)=WF^s(\delta)= \left\{  \begin{array}{cc} \{
(0,\xi) \mid x\in \R\setminus \{ 0\}  \} & s \geq -1/2 \\ \emptyset & s < -1/2
\end{array}\right. 
\end{equation}
and we see how this refines the information in (\ref{verify1}). Like the wave-front set, $WF^s(u)$ is a closed cone in $T^*\R^n\setminus \{0\}$.
Furthermore, part (ii) of propostion \ref{Sobolev prop} entails that $WF^s(u) \subset WF^{s^\prime}(u) \subset WF(u)$ for all $s\leq s^\prime$. In fact, one may show that $WF(u)=\overline{\bigcup_{s\in\R} WF^s(u)}$.
Additionally, if $\varphi \in C^\infty_0 (\R^n)$ does not vanish in a
neighbourhood of $x$ then $(x,\xi)\in WF^s(u)$ if and only if
$(x,\xi)\in WF^s(\varphi u)$; we also have $WF^s(u+w)\subset WF^s(u)\cup
WF^s(w)$. One may show that $WF^s(u)$ can be characterised in a
coordinate-independent way as a subset of the cotangent bundle which
then permits the definition to be extended to distributions on a
manifold by referring back to (any choice of) local coordinates (see,
e.g., remark (i) following Prop.~B.3 in \cite{junker}). 
We shall occasionally use the notation $u\in H^s_\loc(\M)$ if  $WF^s(u)=\emptyset$ for a distribution $u\in\D'(\M)$
(see also the remarks following definition 8.2.5 of \cite{hormander hyper}).

In~\cite{radzikowski}, Radzikowski proved the remarkable result that the
definition of a Hadamard state in terms of the series construction previously given
is equivalent to a condition on the wave-front set of the two-point
function. Namely, the wave-front set is required to lie in a particular
subset of the bicharacteristic set of the Klein--Gordon operator, which
we now define. 

Denote by $\mathcal{R} = \{ (x,\xi)\in T^*\M \, \mid \,
g^{ab}(x)\xi_a\xi_b =0 \, , \, \xi\not=0  \}$ the set of nonzero null covectors over $\M$.
Since $(\M,g)$ is time orientable
we may decompose $\mathcal{R}$ into two disjoint sets $\mathcal{R}^\pm$
defined by $\mathcal{R}^\pm = \{ (x,\xi) \in \mathcal{R} \, \mid \,
\pm\xi \rhd 0 \}$ where by $\xi\rhd 0$ ($\xi\in T^*_x\M$) we mean that $\xi_a$ is in the
dual of the future light cone at $x$. We define the notation
$(x,\xi)\sim (x',\xi')$ to mean that there
exists a null geodesic $\gamma : [0,1]\to \M$ such that $\gamma(0)=x$,
$\gamma(1)=x^\prime$ and $\xi_a = \dot{\gamma}^b(0)g_{ab}(x)$,
$\xi^\prime_a = \dot{\gamma}^b(1)g_{ab}(x^\prime)$. In the instance
where $x=x'$, $(x,\xi)\sim (x,\xi')$ shall mean that $\xi=\xi'$ is null. Then, the set 
\begin{equation}
C = \{ (x,\xi;x',\xi') \in \mathcal{R} \times \mathcal{R}  \mid (x,\xi) \sim (x',\xi')   \}
\end{equation}
is the bicharacteristic relation for the Klein--Gordon operator. We
also define the related sets 
\begin{equation}
C^{+-} = \{ (x,\xi;x^\prime,-\xi^\prime) \in
C \mid  \xi
\rhd 0 \} \, 
\end{equation}
and
\begin{equation}
C^{-+} = \{ (x,-\xi;x^\prime,\xi^\prime) \in
C \mid  \xi\rhd 0 \} \, .
\end{equation}

We may now state the relevant portion of Radzikowski's equivalence theorem~\cite{radzikowski}:

\begin{theorem}\label{rad equiv}
Let $(\M,g)$ be a four-dimensional globally hyperbolic spacetime and
suppose $\Lambda \in \D^\prime (\M\times\M)$ satisifies the Klein--Gordon
equation and has antisymmetric part $iE/2$ modulo smooth functions. Choose a Cauchy
hypersurface $\mathcal{C}$, a causal normal neighbourhood $\mathcal{N}$
of $\mathcal{C}$ and a time function $t$. Then, the following two
conditions are equivalent:
\begin{equation*}
\begin{array}{cl}
i) & \textrm{$\Lambda$ has the Hadamard series structure given by (\ref{hadamard2}) on
$\mathcal{N}\times\mathcal{N}$,} \\
ii) & WF(\Lambda)=C^{+-}.
\end{array}
\end{equation*}
\end{theorem}

As a consequence of this equivalence theorem we may now adopt the
condition that $WF(\Lambda_\omega)=C^{+-}$ as the second \textit{definition} of
the state $\omega$, on $\mathfrak{A}(\M,g)$, being Hadamard.  

Junker and Schrohe ~\cite{junker} applied the theory of Sobolev wave-front sets to
study both Hadamard states and the larger class of adiabatic states (see
\S\ref{adiabaticQEIs}). In particular, lemma 5.2 of \cite{junker} gives the
Sobolev singularity structure of the two point function of Hadamard states.

\begin{theorem}\label{junker schrohe}
Let $\omega$ be a Hadamard state on $\mathfrak{A}(\M,g)$ where $\M$ is a smooth four-dimensional globally hyperbolic spacetime. Then, the two point function, $\Lambda_\omega\in\D^\prime~(~\M~\times~\M~)$, associated to $\omega$ has the following Sobolev wave-front set:
\begin{equation}
WF^s (\Lambda_\omega) = \left\{ \begin{array}{cc} C^{+-} & s \geq -1/2 \\ \emptyset & s<-1/2  \end{array}\right. \, .
\end{equation}
\end{theorem}

\subsection{Sobolev microlocal analysis of the Hadamard series}\label{mlahs}

We will now employ theorem \ref{junker schrohe} to
study the Sobolev wave-front sets of the individual terms in the Hadamard
series, working within an ultra-regular domain $X$ on which the distributions $1/\sigma_+$ and
$\sigma^j\ln\sigma_+$ featuring in (\ref{parametrix}) may be defined. (The length scale $\ell$
will be suppressed from now on.) We shall establish the following statement: 
\begin{equation}
WF^{s+j+1}(\sigma^j\ln\sigma_+)\subset
WF^s(1/\sigma_+) = \left\{ \begin{array}{cc} C^{+-} & s \geq -1/2 \\
\emptyset & s<-1/2    \end{array}  \right. \, .
\end{equation}
In order to do this we observe that the terms appearing in the Hadamard series
are (loosely) related to one another via differentiation. If $P$ is any partial differential operator of
order $r$ on a smooth manifold $\M$, i.e. in local coordinates
\begin{equation}
P = \sum_{|\alpha|\leq r} p_\alpha(x)(-i\partial)^\alpha 
\end{equation} 
where $\alpha$ is a multi-index and $p_\alpha$ are smooth functions, then the \textit{principal symbol}, $p_r(x,\xi)$, of
$P$ is 
\begin{equation}
p_r(x,\xi) = \sum_{|\alpha|=r}p_\alpha(x)\xi^\alpha
\, .
\end{equation}
The \textit{characteristic set} of $P$, $\Char P$, is the set of
$(x,\xi)\in T^*\M\setminus\{ 0\}$ at which the principal symbol
vanishes. Corollaries 8.4.9-10 of \cite{hormander hyper} encapsulate the effect
of partial differential operators on the Sobolev wave-front set of a
distribution:

\begin{lemma}\label{order}
Let $\M$ be a smooth manifold. For $u \in \D^\prime (\M)$ and any
partial differential operator $P$
of order $r$ with smooth coefficients then $WF^s(Pu)\subset WF^{s+r}(u)$ and $WF^{s+r}(u)\subset WF^{s}(Pu)\cup \Char P$.
\end{lemma}

Lemma \ref{order} enables us to quantify our earlier observation about
the relationship between $1/\sigma_+$ and $\sigma^j\ln\sigma_+$:

\begin{prop}\label{order1}
Within an ultra-regular domain we have
\begin{equation} 
WF^{s+1+j}(\sigma^j\ln\sigma_+)\subset
WF^s(1/\sigma_+) \quad \forall s\in\R \,\, \textrm{and} \,\, \forall j \in
\{ 0 \} \cup \N \, .
\end{equation}
\end{prop}
\proof We employ induction on $j$. If $v \in C^\infty (T\M)$ is a smooth vector field, then
$(v\cdot\nabla\otimes\mathbbm{1})\ln\sigma_+ = [(v\cdot\nabla\otimes\mathbbm{1})\sigma]/\sigma_+$. Hence, by lemma
\ref{order}, we have 
\begin{equation}
WF^{s+1}(\ln\sigma_+)\subset WF^s(1/\sigma_+) \cup \Char (v\cdot\nabla\otimes\mathbbm{1}) \, .
\end{equation}
As $v$ is arbitrary,
\begin{equation}
WF^{s+1}(\ln\sigma_+) \subset WF^s(1/\sigma_+)\cup \left( \bigcap_{v\in C^\infty(T\M)}
\Char (v\cdot\nabla\otimes\mathbbm{1}) \right)
\end{equation}
and since 
\begin{equation}
\Char (v\cdot\nabla\otimes\mathbbm{1}) = \{(x,\xi;x^\prime,\xi^\prime)
\in T^* X\setminus 0 \, \mid \, v(x)\cdot\xi =0    \}
\end{equation}
it is clear that the intersection is empty and the statement holds for
$j=0$. Now suppose it holds for some $j\in \{0\}\cup\N$. The identity
\begin{equation}
(v\cdot\nabla\otimes\mathbbm{1})\sigma^{j+1}\ln\sigma_+ =  
[(v\cdot\nabla\otimes\mathbbm{1})\sigma]\left((j+1)\sigma^j\ln\sigma_+ +\sigma^j\right)
\end{equation}
and the inductive hypothesis give
\begin{equation}
WF^{s+2+j}(\sigma^{j+1}\ln\sigma_+)\subset WF^{s+1+j}(\sigma^j\ln\sigma_+)\cup \Char (v\cdot\nabla\otimes\mathbbm{1})
\end{equation}
and taking the intersection over all $v \in C^\infty (T\M)$ as before,
we establish the result for $j+1$ and hence all $j\in\{0\}\cup\N$ by
induction. 
\QED

We next prove the intuitively reasonable result that
$\Lambda_\omega$ is as singular as the leading term in the
Hadamard series.

\begin{prop}\label{easy bit 2}
Let $\omega$ be a Hadamard state. Then, within any ultra-regular domain $X$, we have
\begin{equation}
WF^s(\Lambda_\omega)=WF^s(1/\sigma_+) \qquad \forall s \in
\R\,.
\end{equation}
\end{prop}
\proof
Recall that for every $k\in\N$ there exists a $F_k\in C^k(X)$
such that $\Lambda_\omega =H_k + F_k$. Hence, as
$WF^s(\sigma^j\ln\sigma_+)\subset WF^{s+j+1}(\sigma^j\ln\sigma_+)\subset WF^s(1/\sigma_+)$, 
\begin{eqnarray}
WF^s(\Lambda_\omega) 
&\subset& WF^s(\Delta^\frac{1}{2}/\sigma_+) \cup WF^s(F_k) \, .
\end{eqnarray}
We remark that it is known (from, say, \cite{radzikowski}) that
$\Delta$ does not vanish where $x,x^\prime$ are
null related\footnote{That is, $\Delta^{1/2}$ restricted to $\singsupp
1/\sigma_+$ is non-vanishing.} and as such $WF^s(\Lambda_\omega)\subset
WF^s(1/\sigma_+)\cup WF^s(F_k)$. Moreover, given any particular $s$ we can always find a $k$ sufficiently large
such that $WF^s(F_k)=\emptyset$ and it remains to prove
$WF^s(1/\sigma_+)\subset WF^s(\Lambda_\omega)$. Let $(x,\xi;x',\xi')\in
WF^s(1/\sigma_+)$ such that $WF^{s-\epsilon}(1/\sigma_+)=\emptyset$ for any $\epsilon>0$. Hence, by proposition
\ref{order1},  $WF^{s}(\sigma^j\ln\sigma_+)\subset
WF^{s-j-1}(1/\sigma_+)=\emptyset$ for $j\ge 0$ and we have
$H_k-\Delta^\frac{1}{2}/4\pi^2\sigma_+\in H^s_\loc (X)$. Therefore, $(x,\xi;x',\xi')\in WF^s(H_k)$
and by the nesting property $(x,\xi;x',\xi')\in WF^{s'}(H_k)$ for all
$s'\geq s$. \QED

As a consequence of theorem \ref{junker schrohe} we now
have the Sobolev wave-front sets of the constituent distributions in
the Hadamard series.

\begin{corollary}\label{hadamard orders}
The distributions $1/\sigma_+,\sigma^j\ln\sigma_+ \in \D^\prime
(X)$, where $X$ is an ultra-regular domain, have the following Sobolev wave-front sets:
\begin{eqnarray}
WF^s(1/\sigma_+) &=& \left\{ \begin{array}{cc} C^{+-} & s\geq -1/2 \\ \emptyset & s< -1/2 \end{array}\right. \\
WF^s (\sigma^j \ln\sigma_+) &\subset& \left\{  \begin{array}{cc} C^{+-} & s\geq j+1/2 \\ \emptyset & s<
j
+1/2 \end{array} \right. \,.
\end{eqnarray}
In consequence, we also have, for arbitrary $j\ge -1$, 
\begin{equation}
WF^s(H_j) \subset \left\{ \begin{array}{cc} C^{+-} & s\geq -1/2 \\ \emptyset & s< -1/2
\end{array}\right. 
\label{Hk WFs}
\end{equation}
and
\begin{equation}
WF^s(H_{j+j^\prime}-H_j) \subset \left\{ \begin{array}{cc} C^{+-} & s
\geq j +3/2 \\ \emptyset & s < j +3/2   \end{array} \right. \, \label{Hj diff}
\end{equation} 
for $j'>0$.
\end{corollary}
\proof
As we have already established, $1/\sigma_+$ possesses the lowest order singularity
which lemma \ref{easy bit 2} states is precisely that of $\Lambda_\omega$.
The remaining results follow from proposition~\ref{order1}. \QED

The remainder of this section is devoted to calculating the Sobolev
wave-front sets of the advanced-minus-retarded fundamental
solution $E$ and
a quantity $\widetilde{H}_k\in\D'(X)$ ($X$ an ultra-regular domain) defined by
\begin{equation}
\widetilde{H}_k (x,x^\prime) = \frac{1}{2}\bigg( H_k(x,x^\prime) +
H_k(x^\prime,x) +iE(x,x^\prime)  \bigg) \, ,
\label{Htildek def}
\end{equation}
which plays an important role in our main result theorem \ref{mgi}. As $iE$ is the antisymmetric part of $\Lambda_\omega$, for all Hadamard
$\omega$, theorem \ref{junker schrohe} implies that 
\begin{equation}
WF^s(iE) \subset \left\{ \begin{array}{cc} C^{+-}\bigcup C^{-+} & s \geq -1/2
\\ \emptyset & s<-1/2 \end{array} \right. \, .
\label{WFs E}
\end{equation}

\begin{prop}\label{WFs tilde}
Within an ultra-regular domain, the Sobolev wave-front set of $\widetilde{H}_k$ satisfies
\begin{equation}\label{WFs tilde res}
WF^s(\widetilde{H}_k) \subset \left\{ \begin{array}{cc} C^{+-}\bigcup C^{-+} & s
\geq k+3/2 \\ C^{+-} & s \in [-1/2,k+3/2)\\ \emptyset & s< -1/2  \end{array}
\right. \, .
\end{equation}
\end{prop}
\begin{proof} Suppose first that $s<k+3/2$. It follows from the covariant commutation relations that, 
within an ultra-regular domain $X$, $iE(x,x')=H_{k+2}(x,x')-H_{k+2}(x',x)$ modulo 
$C^{k+2}(X)$. Hence there exists $F\in C^{k+2}(X)$ such that 
\begin{equation} 
[\widetilde{H}_k-H_k](x,x') =
[H_{k+2}-H_k](x,x')-[H_{k+2}-H_k](x',x) + F(x,x') . 
\end{equation} 
As $k+2\ge s$, we have $C^{k+2}(X)\subset H^s_\loc(X)$, so
all three terms on the right-hand side will belong to $H^s_\loc(X)$,
using Eq.~(\ref{Hj diff}) as well. Thus 
$\widetilde{H}_k=H_k$ modulo $H^s_\loc(X)$ for all $s<k+3/2$, which 
establishes (\ref{WFs tilde res}) for $s$ in this range. For $s\geq k+3/2$ the result 
follows from the definition (\ref{Htildek def}) of $\widetilde{H}_k$,
the wave-front set of $H_k$ (and its behaviour under interchange of the
arguments $x$ and $x'$) together with the rule for wave-front sets of
sums of distributions. \end{proof}

Finally, we end this discussion of the microlocal properties with the
following result concerning the singularities of
$\Lambda_\omega-\widetilde{H}_k$ which follows directly from the proof
of corollary \ref{hadamard orders}.

\begin{prop}\label{diff2}
Within an ultra-regular domain, the Sobolev wave-front set of $\Lambda_\omega - \widetilde{H}_k$
is given by
\begin{equation}
WF^s(\Lambda_\omega -\widetilde{H}_k) \subset \left\{ \begin{array}{cc}
C^{+-}\bigcup C^{-+} & s \geq k+3/2 \\ \emptyset & s < k + 3/2  \end{array} \right. \, .
\end{equation}
\end{prop}

\subsection{Restriction results and a point-splitting lemma}

In addition to the results of the previous subsection, our main result
will make use of three additional technical results. The first, Beals'
restriction theorem, enables us to restrict
$\Lambda_\omega, H_k$ and their derivatives to certain submanifolds of $\M\times\M$. The
second, taken from~\cite{fewster}, shows that positive type is preserved under such
restrictions, while the third result is a technical tool that enables us to
write integrals over the diagonal on product manifolds in terms of their
`point-split' Fourier transforms. 

Our QEI results will encompass averages of the stress-energy
tensor smeared over timelike submanifolds, e.g., timelike curves or
hyperplanes, as well as averages over spacetime volumes. For these purposes,
it is necessary to understand how restricting distributions (such as $\Lambda_\omega$,
$H_k$ and their derivatives) to a
submanifold alters the Sobolev wave-front set. A theorem due to Beals
(see Lemma 11.6.1 of \cite{hormander hyper}) tells
us that, for suitably well behaved restrictions, the Sobolev order of
the wave-front set is reduced by an amount proportional to the
codimension of the restriction, while its elements are transformed according
to the associated pull-back mapping. We will state a specialisation of Beals' result to
the case we will need, in which we restrict from a product manifold
$\M\times\M$ to a submanifold $\Sigma\times\Sigma$, where $\Sigma$ is a
submanifold of $\M$. Writing the embedding of $\Sigma$ in $\M$ as a map 
$\iota: \Sigma \rightarrow \M$ and defining 
$\vartheta=\iota\otimes\iota: \Sigma \times\Sigma \to \M \times \M$, 
the restriction of $u\in \D'(\M\times\M)$ to $\Sigma\times\Sigma$ may also
be regarded as the formation of a pull-back $\vartheta^*u$. Beals'
result hinges on the relationship between the Sobolev wave-front set of
$u$ and the conormal bundle $N^*\Sigma$ of $\Sigma$ defined by
\begin{equation}
N^* \Sigma = \{ (\iota(x),\xi)\in T^*\M ; \, x \in \Sigma, \, \iota^*(\xi) = 0 \} \,.
\end{equation}

\begin{theorem}[Beals' restriction theorem]\label{beals}
Let $u \in \D^\prime (\M\times\M)$ and
$\vartheta \in C^\infty (\Sigma\times\Sigma,\M\times\M)$ be defined as
above, and suppose $\M$ and $\Sigma$ have dimensions $m$ and $n$ respectively. If
$\big( N^*\Sigma\times N^*\Sigma\big)\cap WF^s(u)=\emptyset$ for some $s>m-n$ then
$\vartheta^*u$ is a well defined distribution in $\D^\prime(\Sigma\times\Sigma)$. Moreover, 
\begin{equation}
WF^{s-(m-n)}(\vartheta^* u)\subset \vartheta^*WF^s(u)
\end{equation}
where the set $\vartheta^*WF^s(u)$ is defined to be
\begin{eqnarray}
\vartheta^*WF^s (u) &=& \{ (t,\iota^*(\xi);t^\prime,\iota^*(\xi^\prime)) \in
(T^*\Sigma\times T^*\Sigma) \mid \nonumber \\ 
&\,& \qquad (\iota(t),\xi;\iota(t^\prime),\xi^\prime) \in WF^s(u)  \}
\, .
\end{eqnarray}
\end{theorem}

The next result, theorem 2.2 of \cite{fewster}, asserts that the
positive type condition is preserved under the restrictions carried out
by Beals' theorem.

\begin{lemma}\label{res pos} If, in addition to the hypotheses of
Theorem~\ref{beals}, $u\in\D'(\M\times\M)$ is of positive type, then $\vartheta^*u$
is of positive type on $\Sigma\times\Sigma$.
\end{lemma}

Finally, we present a point-splitting identity for distributions of sufficient
regularity. Beginning in $\R^n\times\R^n$, we
have the following. 
\begin{lemma}\label{splitting2}
For all $u\in C_0(\R^n\times\R^n)$, we have the identity
\begin{equation}
\int_{\R^n}\rmd^n t \, u(t,t) = \lim_{\epsilon\to 0^+}
\int_{\R^n}\frac{\rmd^n\xi}{(2\pi)^n}\,e^{-\epsilon|\xi|^2}\widehat{u}(-\xi,\xi) \, .
\end{equation}
In particular, this holds if $u \in H^s (\R^n \times \R^n)\cap\E'(\R^n\times\R^n)$ for
$s>n$, by virtue of the Sobolev embedding of $H^s(\R^n\times\R^n)$ in
$C(\R^n\times\R^n)$. 
\end{lemma}
\proof 
By definition of the Fourier transform, we have
\begin{equation}
\int_{\R^n}\frac{\rmd^n\xi}{(2\pi)^n}\,e^{-\epsilon|\xi|^2}\widehat{u}(-\xi,\xi)
 =\int_{\R^n}\frac{\rmd^n\xi}{(2\pi)^n}\int_{\R^n\times\R^n}
\rmd^n\tau\,\rmd^n\tau'\, e^{-\epsilon |\xi|^2-i\xi\cdot(\tau-\tau')}
u(\tau,\tau')
\end{equation}
As the integrand is absolutely integrable on $\R^{3n}$, Fubini's theorem
permits us to reorder
the integrations and perform the $\xi$ integral first, thus obtaining
\begin{eqnarray}
\int_{\R^n}\frac{\rmd^n\xi}{(2\pi)^n}e^{-\epsilon|\xi|^2}\widehat{u}(-\xi,\xi)
&=& \int_{\R^n\times\R^n}\rmd^n\tau\,\rmd^n\tau'\,\varphi_\epsilon(\tau-\tau') u(\tau,\tau')
\nonumber\\
&=& \int_{\R^n\times\R^n}\rmd^n t\,\rmd^n t'\,\varphi_\epsilon(t') u(t+t'/2,t-t'/2)
\nonumber\\
&=& \int_{\R^n}\rmd^n t'\,\varphi_\epsilon(t') \int_{\R^n}\rmd^n t\, u(t+t'/2,t-t'/2)
\end{eqnarray}
where $\varphi_\epsilon(t) =
(4\pi\epsilon)^{-n/2}e^{-|t|^2/(4\epsilon)}$. We have made the change of variables
$t=(\tau+\tau')/2$, $t'=\tau-\tau'$ (for which the Jacobian is
unity) and reordered
integrals using Fubini's theorem again. As $u$ is continuous and
compactly supported, the inner integral exists for each $t'$ and defines
a continuous compactly supported function. The limit $\epsilon\to 0^+$
exists and yields the value of this function at $t'=0$ because $\varphi_\epsilon$
is an approximate identity. This is the required result. \QED

Note that if $\xi\mapsto \widehat{u}(-\xi,\xi)$ is absolutely integrable on $\R^n$
then the dominated convergence theorem permits us to dispense with the limiting
procedure on the right-hand side.\footnote{Our hypotheses are strong enough to 
guarantee that $\widehat{u}$ is absolutely integrable on
$\R^n\times\R^n$, and hence $\xi\mapsto
\widehat{u}(-\xi+\eta,\xi+\eta)$ is absolutely integrable a.e.\ in
$\eta$ by Fubini's theorem. However, a simple proof of integrability for $\eta=0$ was not forthcoming.}
For our application, we will need a straightforward generalisation of the above
to distributions on manifolds. 
\begin{lemma}\label{splitting3}
Let $(\Sigma,h)$ be a $n$-dimensional pseudo-Riemannian manifold and $u\in
\E'(\Sigma\times\Sigma)\cap H^s(\Sigma\times\Sigma)$ for $s>n$. Suppose
the support of $u$ is contained within $\U\times\U$ where
$\U$ is a single coordinate
chart of $\Sigma$ with associated coordinate map $\kappa : \U \to  \R^n$. Then
\begin{equation}
\int_\Sigma \dvol(x) \, u(x,x) = \lim_{\epsilon\to 0^+}\int_{\R^n}\frac{\rmd^n\xi}{(2\pi)^n} \,e^{-\epsilon|\xi|^2}\widehat{U}(-\xi,\xi)
\end{equation}
where $U:\kappa(\U)\times\kappa(\U)\to \C$ is defined by 
\begin{equation}
U(x,x') =
\big(|h_\kappa|^\frac{1}{4}\otimes |h_\kappa|^\frac{1}{4} \, u_\kappa\big) (x,x')
\, ,
\end{equation}
where $u_\kappa = u\circ (\kappa^{-1}\otimes\kappa^{-1})$ and $h_\kappa$ is
the determinant of the metric in coordinate chart
$\kappa$.
\end{lemma}
\proof
We have
\begin{equation}
\int_\Sigma \dvol(x) u(x,x) = \int_{\R^n} \rmd^n x \, |h_\kappa(x)|^{1/2}
u_\kappa(x,x) = \int_{\R^n} \rmd^n x \, U(x,x)  \, .
\end{equation}
As $h_\kappa$ is a positive smooth function bounded away from zero and
therefore has smooth fractional powers, we may apply lemma \ref{splitting2} to the function
$U$ to obtain the desired result. \QED

\section{An absolute quantum inequality}\label{QEI section}

\subsection{Main result}

We now come to the statement and proof of our main result. Let $\Sigma$
be any $n$-dimensional timelike submanifold of $(\M,g)$ for $1\le n\le
4$, that is, $h=\iota^*g$ is a Lorentzian metric on
$\Sigma$, where $\iota:\Sigma\to \M$ embeds the submanifold $\Sigma$ in
$\M$. We also equip $\Sigma$
with the time orientation induced from $\M$, so that non-zero future-directed causal covectors on
$(\M,g)$ pull back to non-zero future-directed causal covectors on $(\Sigma,h)$.
In our conventions a positive definite metric on a
one-dimensional manifold is regarded as Lorentzian.
As $\Sigma$ is timelike, its tangent space $T\Sigma$ can be annihilated
only by covectors which can annihilate at least one nonzero timelike
vector; in particular, all covectors in the conormal bundle $N^*\Sigma$ are spacelike. 

Our aim is to obtain lower bounds, as $\omega$ varies among Hadamard
states, on quantities of the form
\begin{equation}
\int_\Sigma \dvol(x) \, f^2(x)\big( Q\otimes Q  (\Lambda_\omega - H_2) \big)(x,x)
\end{equation}
where $Q=q^a\nabla_a+b$ is a partial differential operator with smooth real-valued coefficients $q^a$
and $b$ defined on a neighbourhood of $\Sigma$ and $f\in C^\infty_0(\Sigma)$ is real valued.
Note that $\Lambda_\omega-H_2$ is $C^2$ so the coincidence limit is well defined. 
For simplicity it is convenient to assume in addition that $\Sigma$ may
be convered by a single coordinate chart with certain properties.

\begin{defn}
A {\em small sampling domain} is an $n$-dimensional timelike submanifold
$\Sigma$ of $(\M,g)$ such that (i) 
$\Sigma$ is contained in a globally hyperbolic convex normal
neighbourhood in $\M$; (ii) $\Sigma$
may be covered by a single {\em hyperbolic coordinate chart}, i.e.,
a coordinate system $x^0,\dots,x^{n-1}$ on $\Sigma$ with $\partial/\partial x^0$ future-pointing
and timelike, and for which there exists a constant
$c>0$ such that all causal covectors $u_a$ on $\Sigma$ obey
\begin{equation}\label{coord speed of light}
c |u_0| \geq \sqrt{\sum_{j=0}^{n-1} u^2_j}
\end{equation}
(i.e. the coordinate speed of light is bounded from above).
\end{defn}

\noindent A sufficient condition for the existence
of a maximum coordinate speed of light is that $h_{00}>\varepsilon$ and
$|\det(h_{ij})_{i,j=1}^{n-1}|>\varepsilon$  for some $\varepsilon>0$.

It is easy to verify (e.g. by using suitable normal coordinates) that
every point of a general timelike submanifold $\Sigma$ has a neighbourhood
(in $\Sigma$) which is a small sampling domain. Thus any integral over a
compact subset of a timelike submanifold may be decomposed into finitely many
integrals over small sampling domains by a partition of unity. 

Suppose then, that $\Sigma$ is a small sampling domain in $(\M,g)$ with
hyperbolic chart $\{x^a \}_{a=0,\dots,n-1}$. We may express these
coordinates by a map $\kappa : \Sigma\to \R^n$,
$\kappa(p)=(x^0(p),\dots, x^{n-1}(p))$ and write $\Sigma_\kappa =
\kappa(\Sigma)$. Any function $F$ on $\Sigma$ determines a function
$F_\kappa = F\circ \kappa^{-1}$ on $\Sigma_\kappa$; in particular, we
have a smooth map $\iota_\kappa : \Sigma_\kappa \to \M$. The
significance of $\kappa$ being hyperbolic is that the bundle
$\mathcal{R}^+$ of (non-zero) future pointing null covectors on $(\M,g)$ pulls back
under $\iota_\kappa$ so that
\begin{equation}\label{iota R bound}
\iota^*_\kappa\mathcal{R}^+ \subset \Sigma_\kappa \times\Gamma
\end{equation}
where $\Gamma\subset \R^n$ is the set of all $u_a$ with $u_0>0$ and
satisfying (\ref{coord speed of light}), which means that $\Gamma$ is a
proper subset of the upper half-space $\R^+\times\R^{n-1}$ of $\R^n$
(here, we regard $\R^+=[0,\infty)$).

We now state our main result:

\begin{theorem}\label{mgi}
Let $\Sigma$ be a small sampling domain of dimension $n$ in $(\M,g)$
with hyperbolic coordinate map $\kappa$ and suppose $Q$ is a partial
differential operator of order at most one with smooth
real-valued coefficents in a neighbourhood of $\Sigma$. Set
$k=\max\{n+3,5\}$. For any real-valued $f\in
C^\infty_0(\Sigma)$ and any Hadamard state $\omega$ we have the
inequality
\begin{equation}
\int_\Sigma \dvol(x) \, f^2(x)  \big(Q\otimes Q (\Lambda_\omega-H_2) 
\big)(x,x) \geq -\mathcal{B} > -\infty
\end{equation}
where 
\begin{equation}
\mathcal{B} = 2\int_{\R^+\times\R^{n-1}} \frac{\rmd^n\xi}{(2\pi)^n}
\bigg[ |h_\kappa|^\frac{1}{4}f_\kappa \otimes
|h_\kappa|^\frac{1}{4}f_\kappa \, \vartheta^*_\kappa \big( Q\otimes Q \widetilde{H}_k \big)  \bigg]^\wedge
(-\xi,\xi) \, ,
\end{equation}
$h_\kappa$ is the determinant
of the matrix $\kappa^*h$ and $\vartheta  :\Sigma\times\Sigma \to
\M\times\M$ is the map $\vartheta(x,x')=(\iota\otimes\iota)(x,x')$.
\end{theorem}

\noindent\textbf{Remarks}: This bound depends nontrivially on the coordinates (and
on any partition of unity used to reduce a general timelike submanifold into
small sampling domains). In \S\ref{covariance} we will discuss some
classes of QEI averages which, in a sense, determine a natural choice of coordinates.
Note that although $H_2$ is sufficient to renormalise the left-hand
side, the bound is given in terms of $\widetilde{H}_k$ for $k\ge 5$.
Similar results hold when $Q$ has higher order, for suitably
modified values of $k$; we have restricted attention to the cases
relevant to QEIs. 
 
Our strategy will be to mimic the proof of the general worldline quantum
energy inequality presented in \cite{fewster} but with Sobolev wave-front sets,
as opposed to the smooth wave-front sets used in that paper. 

\noindent\textit{Proof of theorem \ref{mgi}.}
We will break the proof into three
parts. Part one will establish that
\begin{eqnarray}\label{part one}
&& \int_\Sigma \dvol(x) \, f^2(x) Q\otimes Q\big( \Lambda_\omega
- H_2 \big)(x,x) \nnnl
&&\!\!\!\!\!\!\!\! = 2\lim_{\epsilon\to 0^+}\int_{\R^+\times\R^{n-1}}\frac{\rmd^n\xi}{(2\pi)^n} \,e^{-\epsilon|\xi|^2}
\bigg[
|h_\kappa|^\frac{1}{4}f_\kappa \otimes |h_\kappa|^\frac{1}{4} f_\kappa
\vartheta^*_\kappa Q\otimes Q \big( \Lambda_\omega - \widetilde{H}_k \big)
\bigg]^\wedge (-\xi,\xi) \, 
\end{eqnarray}
for the given values of $k$. Part two contains a positivity result which enables
us to discard the state dependent contribution to the right hand side of
(\ref{part one}) to obtain the inequality
\begin{eqnarray}
&&  \int_\Sigma \dvol(x) \, f^2(x) Q\otimes Q\big( \Lambda_\omega
- H_2 \big)(x,x)
 \nnnl
&&\!\!\!\!\!\!\!\!  \geq -2 \lim_{\epsilon\to 0^+}\int_{\R^+\times\R^{n-1}}\frac{\rmd^n\xi}{(2\pi)^n} \,e^{-\epsilon|\xi|^2}\bigg[
|h_\kappa|^\frac{1}{4}f_\kappa \otimes |h_\kappa|^\frac{1}{4} f_\kappa
\vartheta^*_\kappa \big( Q\otimes Q \widetilde{H}_k \big)
\bigg]^\wedge (-\xi,\xi) \label{part two}
\end{eqnarray}
Then, in part three, we show that the right-hand side of this expression is finite and equal to the
required lower bound $-\mathcal{B}$.

\noindent \textsc{Part one}: We begin by observing that
$\Lambda_\omega-H_2$ and $\Lambda_\omega-\widetilde{H}_k$ coincide on
the diagonal in $\Sigma\times\Sigma$, so we may write
\begin{eqnarray}
&\,& \int_\Sigma \dvol(x) \, f^2(x) \, \big(Q\otimes Q
(\Lambda_\omega-H_2)\big)(x,x)\nnnl
&\,& \qquad = \int_\Sigma \dvol(x) \, f^2(x) \,
\vartheta^* \big( Q\otimes Q (\Lambda_\omega - \widetilde{H}_k) \big)(x,x).
\end{eqnarray}
The latter form has the merit that $\Lambda_\omega-\widetilde{H}_k$ is symmetric and
more regular than $\Lambda_\omega - H_2$. We have also written in the
restriction map $\vartheta^*$ explicitly, anticipating later steps in
the proof. By hypothesis on $k$, we may
choose $s\in(6,k+3/2)$, and 
Proposition \ref{diff2} tells us that within an ultra-regular domain $X\subset \M\times\M$
\begin{equation}
\Lambda_\omega-\widetilde{H}_k\in
H^{s}_\loc (X)\, .
\end{equation}
As $Q\otimes Q$ is at most second order, lemma \ref{order} entails that
\begin{equation}
Q\otimes Q\big(\Lambda_\omega-\widetilde{H}_k \big) \in
H^{s-2}_\loc (X) \, .
\end{equation}
As the wave-front set $WF^{s-2}(Q\otimes Q(\Lambda_\omega-\widetilde{H}_k))$
is therefore empty and $s-2>4-n$, Beals' restriction theorem, theorem \ref{beals}, yields
\begin{equation}
\vartheta^*Q\otimes Q\big( \Lambda_\omega - \widetilde{H}_k \big)
\in H^{n+s-6}_\loc (\Sigma\times\Sigma) \, ,
\end{equation}
so the point-splitting identity, lemma \ref{splitting3}, may be applied
to give
\begin{eqnarray}\label{step1}
&\, &\int_\Sigma \dvol(x) \, f^2(x) \, \vartheta^*Q\otimes Q \big(\Lambda_\omega- \widetilde{H}_k 
\big)(x,x) \nnnl
&\,& = \lim_{\epsilon\to 0^+}\int_{\R^n}\frac{\rmd^n\xi}{(2\pi)^n}\,e^{-\epsilon|\xi|^2}\bigg[|h_\kappa|^\frac{1}{4}f_\kappa\otimes
|h_\kappa|^\frac{1}{4}f_\kappa\, \vartheta^*_\kappa Q\otimes Q\big(\Lambda_\omega -\widetilde{H}_k  \big) 
\bigg]^\wedge(-\xi,\xi) \, .
\end{eqnarray}
Then, as $\Lambda_\omega-\widetilde{H}_k$ is symmetric and $C^k$, the integrand of
(\ref{step1}) is invariant under $\xi\to-\xi$, so we may replace the integration
over $\R^n$ with that over $\R^+\times\R^{n-1}$ at the expense of a
factor of two, thus obtaining (\ref{part one}).
Note that we are now integrating over those $\xi$ with $\xi_0\ge 0$. This
particular half-space of $\R^n$ is chosen because it contains the cone
$\Gamma$ defined after (\ref{iota R bound}).

\noindent \textsc{Part two:} We now assert that the pull-backs of
$(Q\otimes Q)\Lambda_\omega$ and $(Q\otimes Q)\widetilde{H}_k$ exist
separately, which enables us to split the integrand on the right-hand
side of (\ref{part one}) into two parts. Moreover, we will show that
the first of these, namely $\big[|h_\kappa|^\frac{1}{4} f_\kappa\otimes |h_\kappa|^\frac{1}{4} f_\kappa \,
\vartheta^*_\kappa Q\otimes Q \Lambda_\omega  
\big]^\wedge(-\xi,\xi)$, is nonnegative for all $\xi \in \R^n$. 

The existence of the required pull-backs follows because we have already
observed that non-zero covectors in $N^*\Sigma$ must be spacelike. As
the covectors in the wave-front set of $(Q\otimes Q)\Lambda_\omega$ and $(Q\otimes
Q)\widetilde{H}_k$ are null, it follows that there is no intersection
between their wave-front sets (at any
Sobolev order) and $N^*\Sigma\times N^*\Sigma$ . Therefore the pull-backs exist. 

To establish that $\big[|h_\kappa|^\frac{1}{4} f_\kappa\otimes |h_\kappa|^\frac{1}{4} f_\kappa \,
\vartheta^*_\kappa Q\otimes Q \Lambda_\omega  
\big]^\wedge(-\xi,\xi)\ge 0$, we define
a one parameter family of functions $f_\kappa^\xi(x) = |h_\kappa(x)|^\frac{1}{4}f_\kappa(x)e^{i\xi\cdot x}$ then, as
$f_\kappa \in C^\infty_0(\R^n)$, $[|h_\kappa|^\frac{1}{4}f_\kappa\otimes |h_\kappa|^\frac{1}{4}f_\kappa \,
\vartheta^*_\kappa Q\otimes Q\Lambda_\omega]
\in \E^\prime (\Sigma_\kappa\times\Sigma_\kappa)$ where the Fourier transform is defined
by
$\widehat{u}(\xi,\xi^\prime)=u(e^{i\xi\cdot},e^{i\xi^\prime\cdot^\prime})$
$\forall u \in \E^\prime (\R^n\times\R^n)$. Therefore,
\begin{eqnarray}
&\,& \bigg[ |h_\kappa|^\frac{1}{4}f_\kappa\otimes |h_\kappa|^\frac{1}{4}f_\kappa \,
\vartheta^*_\kappa Q\otimes Q \Lambda_\omega \bigg]^\wedge(-\xi,\xi) \nnnl
&\,& = \bigg[|h_\kappa|^\frac{1}{4}f_\kappa \otimes
|h_\kappa|^\frac{1}{4}f_\kappa\, \vartheta^*_\kappa Q\otimes Q\Lambda_\omega
 \bigg](e^{-i\xi\cdot},e^{i\xi\cdot^\prime}) \\
&\,& = \bigg[\vartheta^*_\kappa Q\otimes Q \Lambda_\omega\bigg](\overline{f^\xi_\kappa},f^\xi_\kappa) \, 
\end{eqnarray}
where we have exploited the fact that $f$ is real valued. It is clear that if $u$ is a distribution of positive type then
$Q\otimes Q u$ is also of positive type because $Q$ has real
coefficients. Accordingly, lemma \ref{res pos} establishes that
$\vartheta^*Q\otimes Q\Lambda_\omega$ is a
distribution of positive type and the assertion is justified. Hence, we
obtain the inequality (\ref{part two}) 
provided that the limit on the right-hand side exists and is finite, which is
the remaining step in the proof. Note that the integral converges for
each $\epsilon>0$ because the Fourier transform of any compactly
supported distribution is polynomially bounded. 

\noindent \textsc{Part three:} Our aim is to show that $I(\xi):=\bigg[|h_\kappa|^\frac{1}{4}f_\kappa\otimes
|h_\kappa|^\frac{1}{4}f_\kappa \,\vartheta^*_\kappa Q\otimes Q \widetilde{H}_k   \bigg]^\wedge(-\xi,\xi)$
is absolutely integrable on the integration region $\R^+\times\R^{n-1}$,
for then we may conclude that the limit on the right-hand side of
(\ref{part two}) exists by dominated convergence and equals $-\mathcal{B}$ (which is thereby
finite). To do this, we introduce an arbitrary Hadamard state $\omega_0$
and use the Hadamard series definition of Hadamard states to write $\widetilde{H}_k=\Lambda_{\omega_0}
+ F_k$ for some $F_k \in C^k(X)$. We consider the contributions of these terms to $I(\xi)$ in
turn. 

First, the results of Radzikowski and Beals entail that
\begin{equation}
WF(\vartheta_\kappa^*Q\otimes Q \Lambda_{\omega_0})\subset
\vartheta_\kappa^*WF(Q\otimes Q \Lambda_{\omega_0})\subset
\vartheta_\kappa^*C^{+-}\subset \vartheta_\kappa^*\left(\mathcal{R}^+\times \mathcal{R}^-\right)
\, ,
\end{equation} 
where $\mathcal{R}^\pm$ are the bundles of future- and past-directed null
covectors defined earlier. Thus, we have
\begin{equation}
WF(\vartheta_\kappa^*Q\otimes Q \Lambda_{\omega_0}) \subset \iota_\kappa^*\mathcal{R}^+\times
\iota_\kappa^*\mathcal{R}^- \subset \left(\Sigma_\kappa\times\Gamma\right)\times\left(\Sigma_\kappa\times (-\Gamma)\right)
\end{equation}
using equation (\ref{iota R bound}) and its obvious analogue for
$\mathcal{R}^-$. 

By Prop.~8.1.3 in \cite{hormander}, it follows that the Fourier transform
of localisations of $\vartheta_\kappa^*Q\otimes Q \Lambda_{\omega_0}$ is
of rapid decay outside the cone $\Gamma\times (-\Gamma)$; in particular we
have rapid decay in the cone $(-\R^+\times\R^{n-1})\times (\R^+\times\R^{n-1})$
(here we have used the assumption that $\Gamma$ is a proper subset of $\R^+\times\R^{n-1}$
because $\kappa$ is hyperbolic). Accordingly we find that 
$[|h_\kappa|^\frac{1}{4}f_\kappa \otimes |h_\kappa|^\frac{1}{4}f_\kappa 
\, \, \vartheta^*_\kappa Q\otimes Q \Lambda_{\omega_0}]^\wedge (-\xi,\xi)$
is rapidly decaying in the integration region $\R^+\times\R^{n-1}$ and
is therefore absolutely integrable there. 

It remains to show that the $F_k$ dependent contribution to $I(\xi)$ is
also absolutely integrable. As $F_k \in C^k(X)$, we have $(Q\otimes Q)F_k\in
C^{k-2}(X)$. Hence there is a constant $c$ such that
\begin{equation}
\bigg|[|h_\kappa|^\frac{1}{4}f_\kappa \otimes
|h_\kappa|^\frac{1}{4}f_\kappa  \, \vartheta^*_\kappa Q\otimes Q F_k]^\wedge(\xi,\xi^\prime)\bigg|
\leq \frac{c}{(1+|\xi|^2+|\xi^\prime|^2)^{(k-2)/2}}
\end{equation}
for all $(\xi,\xi')\in\R^n\times\R^n$ because $f_\kappa$ is compactly supported. As $(k-2)>n$
for the values of $k$ given in the
hypotheses, it follows in particular that left-hand side is absolutely integrable on
$\R^+\times\R^{n-1}$.

Accordingly, we have shown that $I\in L^1(\R^+\times\R^{n-1})$, and the
dominated convergence argument mentioned above completes the proof. \QED

Two points should noted about the foregoing proof. First, the state $\omega_0$ was introduced purely as a
convenient way of showing that our bound is finite; the bound itself
does not depend on any reference state. Second, in the difference QEIs studied
in \cite{fewster} the Gaussian cut-off was not necessary,
because the point-splitting lemma was applied to a smooth compactly supported function.
Moreover, the place of $\widetilde{H}_k$ was taken by the two-point function of a reference state
$\Lambda_{\omega_0}$ and the fact that $WF(\Lambda_{\omega_0})=C^{+-}$
was used to show that the integrand decays rapidly in the integration
region. This line of argument was not available to us here, because
$\widetilde{H}_k$ (in contrast to $H_k$) has a portion of its
wave-front set lying in $C^{-+}$ (see Proposition \ref{WFs tilde}). 

In the next subsection, we will show how Theorem~\ref{mgi} may be used
to obtain QEI bounds, by appropriate choices of the operator $Q$.

\subsection{Examples}

\subsubsection{Worldvolume absolute quantum null energy inequality}\label{example1}

Our first example is a quantum null energy inequality (QNEI), that is, a
lower bound on quantities of the form $\int_\M \dvol\, \langle F^{ab}
T^\mathrm{ren}_{ab}\rangle_\omega$, where $F^{ab}=n^an^b$ and $n^a$ is a smooth, compactly supported null
vector field on $(\M,g)$ that is future-directed where it is nonzero. We will show
how Theorem~\ref{mgi} allows us to obtain an absolute QEI on
$\int_\M \dvol\, \langle F^{ab}
T^\mathrm{ren}_{ab}\rangle_\omega$. To do this, we suppose that $F^{ab}$ is supported within an open subset $\Sigma$ that
is a four-dimensional small sampling domain in $(\M,g)$ with hyperbolic chart $\kappa$.
Noting that
\begin{equation}
\langle F^{ab} T^\mathrm{ren}_{ab}\rangle_\omega (x)= 
\lim_{x^\prime\rightarrow x}\bigg( n^a\nabla_a \otimes n^{b^\prime}\nabla_{b^\prime} \bigg)\big(\Lambda_\omega-H_2
\big)(x,x^\prime) + C_{ab}F^{ab}(x),
\end{equation}
we apply Theorem~\ref{mgi} with $Q=n\cdot\nabla$ and $f\in C_0^\infty(\Sigma)$ chosen to
be real-valued and to equal unity on the support of $F^{ab}$. This
yields the absolute QNEI:
\begin{eqnarray}\label{example1bound}
\int_\M \dvol\, \langle F^{ab}
T^\mathrm{ren}_{ab}\rangle_\omega 
& \geq &  -
2\int_{\R^+\times\R^3}\frac{\rmd^4\xi}{(2\pi)^4} \bigg[|g_\kappa|^\frac{1}{4} \otimes
|g_\kappa|^\frac{1}{4}\left((n\cdot\nabla\otimes n\cdot\nabla) \widetilde{H}_7\right)_\kappa
\bigg]^\wedge(-\xi,\xi)  \nnnl
&\,& \hspace{5mm} + \int_\M \dvol \, F^{ab}C_{ab} \, 
\end{eqnarray}
for all Hadamard states $\omega$. 

Clearly the right-hand side of this inequality depends explicitly on the
choice of coordinates $\kappa$. In Section~\ref{covariance} we will
explain how this problem may be removed by restricting the class of
sampling tensors $F^{ab}$ in such a way that there is a canonical class
of coordinate systems, all of which give the same lower bound.

\subsubsection{Worldline absolute quantum weak energy
inequality}\label{example2}

Our second example applies Theorem~\ref{mgi} to the energy density
sampled along a smooth timelike worldline $\gamma$. This was the
situation studied in \cite{fewster}, where a difference QEI was
obtained. We assume $\gamma$ is given in a proper time parameterisation as a smooth function
$\gamma:I\to\R$, where $I$ is a possibly unbounded open interval of
$\R$ and denote the four-velocity of the curve by $u=\dot{\gamma}$. The
curve forms a small sampling domain, with the proper time
parameterisation as a hyperbolic coordinate system, provided that the
track of $\gamma$ can be contained in a globally hyperbolic convex
normal neighbourhood in $\M$.

The classical energy density of a field $\varphi(x)$ along $\gamma$ may be written in the form
\begin{equation}
u^a u^b T_{ab} (x) = \left(T^\mathrm{split}\big(\varphi\otimes\varphi \big)\right) (x,x)
\end{equation}
where the point split energy density operator is defined within a suitable neighbourhood $\U$ of
$\gamma$ by
\begin{equation}\label{tT}
T^\mathrm{split} = \frac{1}{2}\sum^{3}_{\alpha=0} e^a_\alpha \nabla_a \otimes e^{b^\prime}_\alpha \nabla_{b^\prime} +
\frac{1}{2}\mu^2\mathbbm{1} \otimes\mathbbm{1} \, ,
\end{equation}
and $\{ e^a_\alpha\}_{\alpha=0,1,2,3}$ is any smooth tetrad defined in a
neighbourhood of $\gamma$ such that $e^a_0 = u^a$ on $\gamma$. This
operator may be used to define the renormalised energy density
in the usual fashion. Given any real-valued $f\in C^\infty_0(I)$, we may apply Theorem~\ref{mgi} in turn to
the operators $Q_\alpha = e_\alpha\cdot\nabla$ to obtain 
the absolute quantum weak energy inequality
\begin{eqnarray}
\int_\R \rmd \tau \, f^2(\tau) \langle u^au^b T^\mathrm{ren}_{ab} 
\rangle_\omega(\gamma(\tau)) &\geq & -\int_{\R^+}\frac{\rmd\xi}{\pi} \bigg[
|h_\kappa|^\frac{1}{4}f \otimes |h_\kappa|^\frac{1}{4}f
\vartheta^* T^\mathrm{split}  \widetilde{H}_5  \bigg]^\wedge(-\xi,\xi)
\nnnl
&\,& \hspace{5mm} + \int_\R \rmd \tau \, f^2(\tau) \big( Q + u^au^bC_{ab}
\big)|_{\gamma(\tau)}
\end{eqnarray}
where $\vartheta:(\tau,\tau')\mapsto (\gamma(\tau),\gamma(\tau'))$.

For the purposes of comparison with existing QEI results, let us
consider this bound for the massless Klein--Gordon field in Minkowski spacetime
$(\R^4,\eta)$ for a worldline along the time axis. In this case, the
bound simplifies because the full Hadamard series is
given by the leading term; that is, 
$\Lambda_\omega (x,x') - 1/(4\pi^2\sigma_+(x,x'))$ is smooth and
symmetric for any Hadamard state $\omega$. Of course, $\sigma_+$ is globally defined in
Minkowski space. Now consider the example above, applied to the case
where $\gamma$ is an inertial curve, 
\begin{equation}
\int_\R \rmd \tau \, f^2(\tau) \langle u^au^b T^\mathrm{ren}_{ab} 
\rangle_\omega(\gamma(\tau))
 \geq -\mathcal{B}\stackrel{\mathrm{def}}{=}
-\int_{\R^+} \frac{\rmd\xi}{\pi} \bigg[ f\otimes f \, \vartheta^*T^\mathrm{split}H_{-1}
\bigg]^\wedge (-\xi,\xi)
\end{equation}
where we have denoted $H_{-1}=1/(4\pi^2\sigma_+)=\widetilde{H}_{-1}$. Then,
\begin{eqnarray}
\mathcal{B} &=& \frac{3}{2\pi^2}\int_{\R^+}\frac{\rmd\xi}{\pi}\lim_{\epsilon\rightarrow
0^+}\int_{\R\times\R} \rmd t \, \rmd t' \,
f(t)f(t')\frac{1}{(t-t'-i\epsilon)^4}e^{-i\xi (t-t')} \\
&=&
\frac{3}{2\pi^2}\int_{\R^+}\frac{\rmd\xi}{\pi}\lim_{\epsilon\rightarrow
0^+}\int_{\R} \rmd t \, F(-t) \frac{1}{(t-i\epsilon)^4}e^{-i\xi t}
\end{eqnarray}
where $F(t)=\int_\R \rmd t' \, f(t'-t)f(t')$ has Fourier transform
$\widehat{F}(\xi)=|\hat{f}(\xi)|^2$. Thus
\begin{eqnarray}
\mathcal{B} &=& \frac{1}{4\pi^2}\int_{\R^+}\frac{\rmd \xi}{\pi}\int_{\R^+}
\rmd\zeta |\hat{f}(\xi+\zeta)|^2 \, \zeta^3 \\
&=& \frac{1}{4\pi^2}\int_{\R^+} \rmd\eta \, \int^\eta_0 \rmd\zeta \,
|\hat{f}(\eta)|^2 \zeta^3
\end{eqnarray}
where we have utilised the fact that the Fourier transform of
$1/(t-i0^+)^4$ is $\pi\xi^3\theta(\xi)/3$ \cite{gelfand} and changed
variables to $\eta=\xi+\zeta$. Hence, 
\begin{equation}
\int_\R \, \rmd t \, f^2(t) \langle u^au^b T^\mathrm{ren}_{ab}\rangle_\omega(\gamma(t))
\geq - \frac{1}{16\pi^3}\int_{\R^+} \rmd\eta \, |\hat{f}(\eta)|^2 \eta^4
\end{equation}
which is the same as the QEI for the massless field in Minkowski
spacetime obtained in \cite{fewster eveson}.

This example is of particular importance as on small length
scales one expects the massive quantum field in a curved background to behave like
its massless counterpart in flat spacetime. We expect that the same should hold for the quantum
inequalities, i.e., on small length scales the dominant contribution
to the bound arises from the $1/\sigma_+$ contribution to the Hadamard
series. This will be investigated in a future work.

\subsubsection{QEIs for adiabatic states}\label{adiabaticQEIs}

Finally, we show how our analysis of the Hadamard series using the Sobolev
wave-front set allows us to establish QEIs for adiabatic
states. We refer the reader to \cite{junker} for a detailed study of adiabatic states and
further references. Adiabatic states, like Hadamard states, are defined in terms of
their singular structure: following
\cite{junker},\footnote{In~\cite{junker} the definition of adiabatic
states was given, as for Hadamard states, only for quasi-free states, so
the present usage is a slight extension.} a state
$\omega$ on $\mathfrak{A}(\M,g)$ is an \textit{adiabatic state of order
$N$} if its associated two point function $\Lambda_\omega$ satisfies
$WF^s(\Lambda_\omega)\subset C^{+-}$ for all $s< N+3/2$. From this definition, we
see that any Hadamard state is adiabatic to all orders. In what follows
the following lemma, taken from \cite{junker}, is essential:

\begin{lemma}\label{adiabatic}
Let $\omega$ be a Hadamard state and $\omega'$ be an adiabatic state of
order $N$, with associated two point functions
$\Lambda_\omega,\Lambda_{\omega'}$ respectively, on $\mathfrak{A}(\M,g)$. Then
\begin{equation}
WF^s(\Lambda_\omega - \Lambda_{\omega'}) = \emptyset \label{adiabatic difference}
\end{equation}
for all $s<N+3/2$.
\end{lemma}

An immediate corollary is that (\ref{adiabatic difference}) also holds for all
$s<N+3/2$ if $\omega$ and $\omega'$ are any two adiabatic states of
order $N$. By the Sobolev embedding theorem (Proposition \ref{Sobolev prop}
(i)), differences of
this type will be in $C^2(\M\times\M)$ provided $N>9/2$, thus permitting
the construction of a normal ordered stress-energy tensor. Similarly, if
$\omega$ is adiabatic of order $N>9/2$ and $k\ge 2$, a difference of the
form $\Lambda_\omega-H_k$ will be twice continuously differentiable on an
ultra-regular domain, permitting the computation of the renormalised
stress-energy tensor. It is straightforward to modify the proof of
Theorem~\ref{mgi} to obtain the following. 

\begin{theorem} (a) Theorem~\ref{mgi} continues to hold (with the same lower bound)
under the weaker hypothesis that $\omega$
is an adiabatic state of order $N>9/2$. (b) Using the assumptions and
notation of Theorem~\ref{mgi}, except that $\omega$ is assumed only to
be an adiabatic state of order $N>9/2$, there is a difference inequality
\begin{eqnarray}\label{adiabatic ineq}
&\,& \int_\Sigma \dvol(x) f^2(x) \, Q\otimes Q \big( \Lambda_\omega-\Lambda_{\omega'}\big)(x,x) \nnnl
&\,& \quad \geq - 2\int_{\R^+\times\R^{n-1}}\frac{\rmd^n\xi}{(2\pi)^n} 
\bigg[ |h_\kappa|^\frac{1}{4}f_\kappa \otimes |h_\kappa|^\frac{1}{4}f_\kappa \,\vartheta^*_\kappa Q\otimes Q \Lambda_{\omega'}  \bigg]^\wedge(-\xi,\xi) \, ,
\end{eqnarray}
for any reference state $\omega'$ which is adiabatic of order $N'>n+11/2$.
\end{theorem}
\begin{proof}
We sketch the main points only. For (a), note that the hypotheses on $N$
and $k$ entail that we may choose $s\in(6,\min\{N,k\}+3/2)$. Part one of
the proof of Theorem~\ref{mgi} will continue to hold provided that
$\Lambda_\omega-\widetilde{H}_k\in H^s_{\rm loc}$. To see this, we
introduce an arbitrary Hadamard state $\omega_0$ and note that
\begin{equation}
WF^s(\Lambda_\omega-\widetilde{H}_k)\subset
WF^s(\Lambda_\omega-\Lambda_{\omega_0})\cup WF^s(\Lambda_{\omega_0}-\widetilde{H}_k)
=\emptyset
\end{equation}
using Lemma~\ref{adiabatic} together with the fact that $s<N+3/2$, and Proposition
\ref{diff2} together with $s<k+3/2$. Part two of the proof holds because
$\Lambda_\omega$ is a bisolution to the Klein--Gordon equation, so all
covectors in its wave-front set are null, from which it follows that the
required pull-back exists. The third part is identical to the original
argument. 
For (b), the hypotheses on $N$ and $N'$ permit us to choose
$s\in(6,\min\{N,N'\}+3/2)$, and we have
$WF^s(\Lambda_\omega-\Lambda_{\omega'})=\emptyset$ by the remark
following Lemma~\ref{adiabatic}. Part one of the argument then goes
through, as does the second part (with $\Lambda_{\omega'}$ replacing
$\widetilde{H}_k$). The remaining issue is to check that the bound is
finite. Introducing a reference Hadamard state $\omega_0$ as before, we
note that $\Lambda_{\omega'}-\Lambda_{\omega_0}\in H^{s'}_{\rm
loc}(\M\times\M)$ for some $s'\in(n+7,N'+3/2)$, and hence
$\Lambda_{\omega'}-\Lambda_{\omega_0}\in C^{n+3}(\M\times\M)$ (this is
the reason for the constraint $N'>n+11/2$). This is sufficient for Part
three to apply to $\Lambda_{\omega'}$ in place of $\widetilde{H}_k$.
\end{proof}

\subsection{Covariance}\label{covariance}

The QEIs obtained from theorem \ref{mgi} are not covariant in full
generality because they 
depend non-trivially on the coordinates used, and, in some cases, on a
choice of tetrad near $\Sigma$. However, covariance may be rescued if we can restrict the
freedom to choose coordinates and the tetrad in a covariant
fashion so that the bound is independent of any residual choice. This
strategy was successfully employed in \cite{fewster
pfenning 2} for worldline difference QEIs, and the same techniques would also
apply to our worldline bounds; here we show how this may be
accomplished for worldvolume averages (timelike submanifolds of other
dimensions could be handled in an analogous fashion, but we will not do
this here for brevity). 

Consider the quantum null energy inequality studied in section~\ref{example1}. We will show that if 
the sampling tensor $F^{ab}$ picks out a preferred smooth timelike curve $\gamma$ in a covariant way
and we employ a system of Fermi normal coordinates near $\gamma$, then residual choices in our
construction cannot affect the bound. With these restrictions, our
absolute QEI would be locally covariant in the sense of \cite{fewster pfenning
2}; see also \cite{fewster loc cov} for a more abstract discussion of
these ideas in the formulation of locally covariant quantum field theory
developed by Brunetti,
Fredenhagen \& Verch \cite{brunetti} in terms of category theory. 

The requirement that $F^{ab}$ should pick out a unique timelike curve may be addressed in various
ways. For example, if we restrict
to sampling tensors for which there exists a (necessarily unique) pair of points $x,x' \in \M$ such
that the support of the sampling tensor obeys
\begin{equation}
\supp F^{ab} = J^-(x)\cap J^+(x')
\end{equation}
and is contained within a convex normal neighbourhood then the unique timelike
geodesic between $x'$ and $x$ may be used as our choice of $\gamma$.
From now on we assume that the sampling tensor does indeed select a
preferred timelike curve, and that $\gamma$ is given in a proper time
parameterisation.

As already mentioned, we will restrict our coordinate system to belong
to the class of Fermi normal coordinates about $\gamma$. For
completeness, we briefly summarise the salient features
of Fermi--Walker transport and Fermi normal coordinates, mainly following chapters 1 \S4 and 2 \S10 of \cite{synge}.
Recall that a vector field $\xi$ defined on $\gamma$ is said to be \textit{Fermi--Walker transported} along it if
$D_{FW}\xi=0$, where
\begin{equation}
D_{FW}\xi^a = (\dot{\gamma}\cdot\nabla)\xi^a -
g_{bc}\big(\dot\gamma^c \alpha^a-\alpha^c\dot\gamma^a  \big)\xi^b
\end{equation}
and $\alpha^a=\dot\gamma\cdot\nabla\dot\gamma^a$. Since
$\alpha\cdot\dot\gamma=0$ and $\dot\gamma^2=1$ it is easy to see that
$D_{FW}\dot\gamma=0$; hence, the velocity vector is preserved
under Fermi--Walker transport. Moreover, it is possible to show that
Fermi--Walker transport of two vectors along $\gamma$ preserves their
inner-product. Therefore, a tetrad remains an orthogonal
frame along $\gamma$ under Fermi--Walker transport. If $\gamma$ is a
timelike geodesic, then Fermi--Walker and parallel transport coincide. 

The construction of Fermi normal coordinates near $\gamma$ proceeds
as follows. Let $y$ lie on $\gamma$ and construct
an oriented and time-oriented orthonormal frame $\{ e^a_\alpha \}_{\alpha=0,1,2,3}$ at $y$ with $e^a_0 =
\dot{\gamma}^a|_y$. Fermi--Walker transport yields a tetrad along the
whole of $\gamma$. In a convex normal neighbourhood $\U$ of $\gamma$ 
each point $x\in\U$ will be joined to $\gamma$ by a unique
spacelike geodesic segment $c$ which is orthogonal to $\gamma$ and which meets
it at some $\gamma(t)$. Assuming that $c$ is parameterised by proper
length, the Fermi normal coordinates $x^a$ of $x$ are
\begin{equation}
x^0=t; \qquad x^i=s\dot{c}\cdot e_i|_{\gamma(t)}\,,
\end{equation}
where $s$ is the proper length of $c$. 

This construction has two important features. First, the metric takes the Minkowski form in these
coordinates everywhere on $\gamma$. By continuity, this guarantees that the Fermi
normal coordinates form a hyperbolic chart in a neighbourhood of
$\gamma$. Second, the only freedom in the construction is the choice of origin on $\gamma$ (which amounts to the freedom to
add a constant to $x^0$) and the choice of the spatial tetrad vectors $e_i$
($i=1,2,3$) at $y$, which are determined only up to a rotation. Owing to the angle-preserving nature
of Fermi--Walker transport, any two coordinate systems obtained by the
construction are therefore globally related by ${x'}^0=x^0+\lambda$,
${x'}^i=S^i_{\phantom{i}j}x^j$ for constant scalar $\lambda$ and constant rotation matrix
$S\in SO(3)$. 

If that the sampling tensor is supported within the neighbourhood
of $\gamma$ in which the Fermi normal coordinates are hyperbolic, it is
easy to see that the absolute QEI (\ref{example1bound}) is independent of the particular system of
Fermi normal coordinates chosen. The key point is that the Jacobian determinant for a change of
coordinates between two Fermi normal coordinate systems is identically
unity by the remarks given above. (Note also that $F^{ab}$ may be written
uniquely as $F^{ab}=n^an^b$ under the constraint that $n^a$ is
future-pointing and null where it is nonzero.)

For more general QEIs one also needs to construct a tetrad throughout
the support of the sampling tensor. This may be done by taking the
tetrad formed along $\gamma$ and propagating it by parallel transport along spacelike
geodesics which meet $\gamma$ orthogonally. 

\section{Conclusion}

We have given the first explicit absolute quantum energy
inequalities for the massive minimally coupled Klein--Gordon field in arbitrary
four-dimensional globally hyperbolic backgrounds, by refining the
argument of~\cite{fewster} to make use of the theory of the Sobolev wave-front
set, and analysing microlocal properties of the components of the Hadamard series. The lower bounds
are given in terms of partial sums of the Hadamard series, which are
computed locally. Previously explicit absolute quantum energy
inequalities were known only for the massless field in two
dimensions~\cite{flanagan2} (a similar argument could be used to extend
this to general positive energy unitary conformal field theories, based
on~\cite{fewster hollands}). Although the
bounds make use of coordinate systems, we have shown that by restricting
the class of sampling tensors, there are circumstances in which the bound is covariant. 

Absolute QEIs may also be found for higher spin fields. In the case of the Dirac
field, which will be reported elsewhere \cite{smith}, one adapts the
difference QEI obtained in \cite{dawson fewster} in a similar fashion to
the way in which~\cite{fewster} has been adapted here. Moreover, it is expected that
one should be able to employ our
method to prove an absolute QEI for the spin-1 vector bosons, using the
formulation of the Hadamard condition for the Maxwell and Proca fields
given in \cite{fewster pfenning}.

One possibility which is opened up by our work is to obtain control over
the size of spacetime region in which the absolute QEI bound can be
approximated to a good degree by the QEIs obtained in Minkowski space
for massless fields. This would potentially result in very simple bounds
of wide applicability, and is the subject of ongoing work.

\begin{appendix}

\section{Hadamard recursion relations}\label{hadamard recursion relations}

In this appendix we briefly summarise the method for generating the
coefficient functions, $\{ v_j \}_{j=0,\dots,k}$ and $\{
w_j\}_{j=0,\dots,k}$, featuring in (\ref{parametrix}). These are
obtained as the coefficients in the formal power series solution
\begin{eqnarray}
H(x,x^\prime) &=&  \frac{1}{4\pi^2} \bigg\{ \frac{\Delta^\frac{1}{2}(x,x^\prime)}{\sigma_+
(x,x^\prime)} + \sum^\infty_{j=0}v_j
(x,x^\prime)\frac{\sigma^j(x,x^\prime)}{\ell ^{2(j+1)}}\ln\bigg(\frac{\sigma_+ (x,x^\prime)}{\ell^2}\bigg)
\nonumber \\
&\,& \qquad 
+\sum^\infty_{j=0}w_j(x,x^\prime)\frac{\sigma^j(x,x^\prime)}{\ell^{2(j+1)}}\bigg\}
\, .
\end{eqnarray}
to 
\begin{equation}
\big( (\nabla^2 + \mu^2) \otimes\mathbbm{1}\big) H_k(x,x^\prime) =0 \quad \textrm{subject
to}\quad 
 w_0 = 0 \, .
\end{equation}
The series does not actually converge except in analytic spacetimes,
which is why one makes use of the partial sums $H_k$. 
The recursion relations for the $v_j$ for the massive field in a curved background are:
\begin{eqnarray}
0 &=& \ell^2(\nabla^2+\mu^2)\Delta^\frac{1}{2} + 2\nabla v_0 \cdot \nabla\sigma + 4v_0
+v_0\nabla^2\sigma \label{v_0}\\
0 &=& \ell^2(\nabla^2+\mu^2)v_j + 2(j+1)\nabla v_{j+1}\cdot\nabla\sigma
\nnnl
&\,& \quad - 4j(j+1)v_{j+1} + (j+1)v_{j+1}\nabla^2\sigma \label{v_j}
\end{eqnarray}
where $j\in \{ 0 \} \cup \N$. In a regular domain $X$ the system of differential
equations uniquely determines the series of $v_j$'s.
The $w_j$ series is specified once the value of $w_0$ is
fixed; we have adopted Wald's prescription that $w_0=0$ \cite{wald} and with this
boundary condition the recursion relations are:
\begin{eqnarray}
0 &=& 2\nabla w_1\cdot\nabla\sigma + w_1\nabla^2\sigma + 2\nabla v_1
\cdot \nabla\sigma - 4v_1 + v_1\nabla^2\sigma \label{w_1}\\
0 &=& \ell^2(\nabla^2+\mu^2)w_{k} +2(k+1)\nabla w_{k+1}\cdot
\nabla\sigma \nnnl 
&\,& \quad - 4k(k+1)w_{k+1} + (k+1)w_{k+1}\nabla^2\sigma \nnnl
&\,& \quad +2\nabla v_{k+1}\cdot \nabla\sigma -4(2k+1)v_{k+1} +
v_{k+1}\nabla^2\sigma \label{w_k}
\end{eqnarray}
where $k\in\N$. The system of equations
(\ref{v_0},\ref{v_j},\ref{w_1},\ref{w_k}) are known as the Hadamard
recursion relations; these relations for the massless field may be found
in \cite{adler, dewitt} where the dependency on a choice of length scale
$\ell$ is suppressed.

\end{appendix}

\begin{center}
\textbf{Acknowledgements}
\end{center}
The authors would like to thank S.P. Dawson and L. Osterbrink for their many helpful comments on the manuscript, and J. Schlemmer and R. Verch for pointing out some typographical errors in the Appendix.


\begin{thebibliography}{99}
\renewcommand{\itemsep}{-2pt}

\bibitem{adler}
Adler S.L., Lieberman J. \& Ng Y.J., 1976, \textit{Regularization of the stress energy tensor for vector and scalar particles propagating in a general background metric}, Ann. Phys. \textbf{106}, 279-321

\bibitem{alcubierre}
Alcubierre M., 1994, \textit{The warp drive: hyper-fast travel within
general relativity}, Class. Quantum Grav. \textbf{11}, L73-L77 

\bibitem{brunetti fredenhagen} 
Brunetti R. \& Fredenhagen K., 2000, \textit{Microlocal Analysis and Interacting
Quantum Field Theories: renormalization on Physical Backgrounds},
Commun. Math. Phys. \textbf{208}, 623-661

\bibitem{brunetti}
Brunetti R., Fredenhagen K. \& Verch R., 2003, \textit{The generally
covariant locality principle - A new paradigm for local quantum field
theory}, Commun. Math. Phys. \textbf{237}, 31-68

\bibitem{dawson}
Dawson S.P., 2006, \textit{A quantum weak energy inequality for the Dirac
field in two-dimensional flat spacetime}, Class. Quantum Grav.
\textbf{23}, 287-293

\bibitem{dawson fewster}
Dawson S.P. \& Fewster C.J., 2006, \textit{An explicit quantum weak energy inequality for Dirac fields in curved
spacetimes}, Class. Quantum Grav.
\textbf{23}, 6659-6681

\bibitem{dewitt}
DeWitt B.S. \& Brehme R.W., 1960, \textit{Radiation damping in a gravitational field}, Ann. Phys. \textbf{9}, 220-259

\bibitem{epstein}
Epstein H., Jaffe A. \& Glaser V., 1965, \textit{Nonpositivity in the
energy density in quantised field theories}, Nuovo Cimento \textbf{36}, 1016-1022

\bibitem{fewster eveson}
Fewster C.J. \& Eveson S.P., 1998, \textit{Bounds on negative energy densities in flat spacetime}, Phys. Rev. D \textbf{58}, 084010

\bibitem{fewster}
Fewster C.J., 2000, \textit{A general worldline quantum inequality}, Class. Quantum Grav. \textbf{17}, 1897-1911

\bibitem{fewster teo}
Fewster C.J. \& Teo E., 1999, \textit{Bounds on negative energy densities in static space-times},
Phys. Rev D \textbf{59}, 104016

\bibitem{fewster verch}
Fewster C.J. \& Verch R., 2002, \textit{A quantum weak energy inequality
for Dirac fields in curved spacetime}, Commun. Math. Phys. \textbf{225}, 331-359

\bibitem{fewster mistry}
Fewster C.J. \& Mistry B., 2003, \textit{Quantum weak energy
inequalities for the Dirac field in flat spacetime}, Phys. Rev. D
\textbf{68}, 105010

\bibitem{fewster pfenning}
Fewster C.J. \& Pfenning M.J., 2003, \textit{A weak quantum energy inequality for spin-one fields in curved
spacetime}, J. Math. Phys. \textbf{44}, 4480-4513

\bibitem{fewster 2d}
Fewster C.J., 2002, \textit{Quantum energy inequalities in two dimensions}, Phys. Rev. D
\textbf{70}, 127501

\bibitem{fewster lisbon}
Fewster C.J., 2005, \textit{Energy inequalities in quantum field theory},
in {\it XIVth International Congress on Mathematical Physics}, ed. J.C. Zambrini (World Scientific, Singapore, 2005).
See {\tt math-ph/0501073} for an expanded and updated version.

\bibitem{fewster paris}
Fewster C.J., 2005, \textit{Quantum energy inequalities and stability
conditions in quantum field theory}, in {\em Rigorous Quantum Field
Theory: A Festschrift for Jacques Bros}, A.~Boutet de Monvel,
D.~Buchholz, D.~Iagolnitzer, U.~Moschella (Eds.) Progress in
Mathematics, Vol. 251. (Birkh\"auser, Boston, 2007), {\tt math-ph/0502002}. 

\bibitem{fewster hollands}
Fewster C.J. \& Hollands S., 2005, \textit{Quantum energy inequalities in
two-dimensional conformal field theory}, Rev. Math. Phys. \textbf{17}, 577-612

\bibitem{fewster roman}
Fewster C.J. \& Roman T.A., 2005, \textit{On wormholes with arbitrarily
small quantities of exotic matter}, Phys. Rev. D \textbf{72}, 044023

\bibitem{fewster pfenning 2}
Fewster C.J. \& Pfenning M.J., 2006, \textit{Quantum energy inequalities and
local covariance I: Globally hyperbolic spacetimes}, J. Math. Phys. \textbf{47}, 082303

\bibitem{fewster osterbrink}
Fewster C.J. \& Osterbrink L.W., 2006, \textit{Averaged inequalities for
the non-minimally coupled classical scalar field}, Phys. Rev. D
\textbf{74}, 044021

\bibitem{fewster osterbrink II}
Fewster C.J. \& Osterbrink L.W., 2007,
\textit{Quantum Energy Inequalities for the Non-Minimally Coupled Scalar
Field}, J. Phys. A:Math. Theor. \textbf{41}, 025402

\bibitem{fewster loc cov}
Fewster C.J., 2006, \textit{Quantum energy inequalities and local covariance II: Categorical
formulation}, Gen. Rel. and Grav. \textbf{39}, 1855-1890 

\bibitem{flanagan1}
Flanagan \'E.\'E., 1997, \textit{Quantum inequalities in two-dimensional
Minkowski spacetime}, Phys. Rev. D \textbf{56}, 4922-4926

\bibitem{flanagan2}
Flanagan \'E.\'E., 2002, \textit{Quantum inequalities in two-dimensional
curved spacetimes}, Phys. Rev. D \textbf{66}, 104007

\bibitem{ford}
Ford L.H., 1978, \textit{Quantum coherence effects and the second law of
thermodynamics}, Proc. R. Soc. Lond. A. \textbf{364}, 227-236

\bibitem{ford and roman}
Ford L.H. \& Roman T.A., 1996, \textit{Quantum field theory constrains
traversable wormhole geometries}, Phys. Rev. D \textbf{53}, 5496-5507

\bibitem{ford and pfenning}
Ford L.H. \& Pfenning M.J., 1997, \textit{The unphysical nature of ``warp
drive"}, Class. Quantum Grav. \textbf{14}, 1743-1751

\bibitem{fulling sweeny wald} 
Fulling S.A., Sweeny M. \& Wald R.M., 1978, \textit{Singularity structure of
the two-point function in quantum field theory in curved spacetime},
Commun. Math. Phys. \textbf{65}, 257-264

\bibitem{fulling narcowich wald}
Fulling S.A., Narcowich F.J. \& Wald R.M., 1981, \textit{Singularity structure
of the two-point function in quantum field theory in curved spacetime
II}, Ann. Phys. (N.Y.) \textbf{136}, 243-272

\bibitem{gelfand}
Gel'fand I.M. \& Shilov G.E., 1964, \textit{Generalised functions}, Academic Press, New York and London

\bibitem{gunther}
G\"unther P., 1988, \textit{Huygen's principle and hyperbolic equations},
Academic Press Inc, New York

\bibitem{hollands wald a} 
Hollands S. \& Wald R.M., 2001, \textit{Local Wick Polynomials and Time
Ordered Products of Quantum Fields in Curved Spacetime},
Commun. Math. Phys. \textbf{223}, 289-326 

\bibitem{hollands wald b} 
Hollands S. \& Wald R.M., 2002, \textit{Existence of Local Covariant
Time Ordered Products of Quantum Fields in Curved Spacetime},
Commun. Math. Phys. \textbf{231}, 309-345

\bibitem{hormander}
H\"ormander L., 1989, \textit{The analysis of linear partial differential operators I}, second
edition, Springer-Verlag, New York

\bibitem{hormander hyper}
H\"ormander L., 1996, \textit{Lectures on nonlinear hyperbolic differential equations}, Springer, New York

\bibitem{hu ling zhang}
Hu B., Ling Y. \& Zhang H., 2006, \textit{Quantum inequalities for massless
spin-3/2 field in Minkowski spacetime}, Phys. Rev. D \textbf{73}, 045015

\bibitem{junker}
Junker W. \& Schrohe E., 2002, \textit{Adiabatic vacuum states on general spacetime manifolds: Definition, construction, and physical
properties}, Annales Poincar\'{e} Phys. Theor. \textbf{3}, 1113-1182

\bibitem{kay}
Kay B.S. \& Wald R.M., 1991, \textit{Theorems on the uniqueness and thermal properties
of stationary, nonsingular, quasifree states on spacetimes with a bifurcate Killing horizon}, Phys. Rep. \textbf{207}, 49-136

\bibitem{moretti}
Moretti V., 2003, \textit{Comments on the stress-energy tensor operator
in curved spacetime}, Commun. Math. Phys. \textbf{232}, 189-221

\bibitem{pfenning}
Pfenning M., 1998, PhD Thesis \textit{Quantum inequality restrictions on negative
energy densities on curved spacetimes}, pre-print {\tt gr-qc/9805037}

\bibitem{radzikowski}
Radzikowski M., 1996, \textit{Micro-local approach to the Hadamard condition in quantum field theory on curved
space-time}, Commun. Math. Phys. \textbf{179}, 529-553

\bibitem{reed2}
Reed M. \& Simon B., 1975, \textit{Methods of modern mathematical physics, Vol II, Fourier analysis and self adjointness}, Academic Press, New York

\bibitem{sahlmann2}
Sahlmann H. \& Verch R., 2001, \textit{Microlocal spectral condition and
Hadamard form for vector-valued quantum fields in curved spacetime},
Rev. Math. Phys. \textbf{13}, 1203-1246

\bibitem{smith}
Smith C.J., 2007, \textit{An absolute quantum energy inequality for the
Dirac field in curved spacetime}, Class. Quantum Grav. \textbf{24}, 4733-4750

\bibitem{synge}
Synge J., 1960, \textit{Relativity: The general theory}, North Holland, Amsterdam

\bibitem{vollick}
Vollick D.N., 2000, \textit{Quantum inequalities in curved two dimensional
spacetimes}, Phys. Rev. D \textbf{61}, 084022

\bibitem{wald}
Wald R.M., 1978, \textit{Trace anomaly of a conformally invariant quantum field in curved spacetime}, Phys. Rev. D \textbf{17}, 1477-1484  

\bibitem{yu wu}
Yu H. \& Wu P., 2004, \textit{Quantum inequalities for the free Rarita-Schwinger fields in flat spacetime}, Phys. Rev. D \textbf{69}, 064008

\end{thebibliography}
\end{document}